\documentclass[conference]{IEEEtran}
\IEEEoverridecommandlockouts

\usepackage{cite}
\usepackage{amsmath,amssymb,amsfonts}
\usepackage{amsthm}
\usepackage{graphicx}

\usepackage{textcomp}
\usepackage{xcolor}
\usepackage{blindtext}
\usepackage[normalem]{ulem}
\usepackage{graphicx} 
\usepackage{url}
\usepackage{mathtools}
\usepackage{capt-of} 

\usepackage{xcolor} 

\usepackage[linesnumbered,ruled,vlined]{algorithm2e}
\usepackage[hidelinks]{hyperref} 
\usepackage{orcidlink}
\usepackage{amsmath}
\usepackage{booktabs}
\usepackage{colortbl}
\usepackage{enumerate}
\usepackage[mathscr]{eucal}
\usepackage{graphicx}
\usepackage{makecell}
\usepackage{mathtools}
\usepackage[font=rm]{caption}
\DeclareCaptionType{copyrightbox}
\usepackage{shortvrb}
\usepackage{tabularx}
\usepackage{threeparttable}
\usepackage{verbatim}
\usepackage{xspace}
\usepackage{caption}
\usepackage{subcaption}
\usepackage{listings}
\usepackage{fancyvrb}
\usepackage{enumitem}
\usepackage{multirow}
\usepackage{amsmath}
\usepackage{bbding}
\usepackage{siunitx}
\usepackage{bm}
\usepackage{algpseudocode}
\usepackage{textcomp}
\usepackage{xparse}
\setlist[itemize,1]{leftmargin=3mm,itemsep=0mm}
\setlist[enumerate,1]{leftmargin=5mm,itemsep=0mm}
\usepackage{balance}
\usepackage{xspace}
\usepackage{ragged2e}
\newif\ifcameraready
\camerareadyfalse

\usepackage{marginnote}

\usepackage{array}
\usepackage{pifont}
\usepackage{tikz}
\usepackage[most]{tcolorbox}
\numberwithin{equation}{section}
\def\BibTeX{{\rm B\kern-.05em{\sc i\kern-.025em b}\kern-.08em
    T\kern-.1667em\lower.7ex\hbox{E}\kern-.125emX}}

\newcommand{\feng}[1]{{\color{purple}{\bf{Feng says:}} \emph{#1}}}

\newtcolorbox{ExampleBox}[1][]{%
  colback=black!6,
  colframe=black!35,
  boxrule=0.4pt,
  arc=1mm,
  left=1.2mm,right=1.2mm,top=0.9mm,bottom=0.9mm,
  fontupper=\normalfont,
  #1
}
\newtheoremstyle{exstyle}%
  {1pt}   
  {1pt}   
  {\itshape}{}%
  {\itshape}{.}%
  {0.5em}{}%
\theoremstyle{exstyle}
\newtheorem{ex}{Example}

\newcommand{\update}[1]{{\color{black}#1}}

\newtheorem{defn}{Definition}
\newtheorem{cdefn}{Concept}
\usepackage{amsthm}
\newcommand{\cmark}{\ding{51}} 
\newcommand{\xmark}{\ding{55}} 
\newcommand{\ourmethod}{\textsf{SemiBonsai}}
\newcommand{\router}{{Budget-Aware Bandit Router}}
\newcommand{\qsubset}{\textsc{Complex-Q}}
\newcommand{\tsubset}{\textsc{Complex-T}}
\newcommand{\qtsubset}{\textsc{Complex-QT}}

\newcommand{\tablestructurer}{Multiway Layered Table Structurer}
\newcommand{\uncertaintyresolver}{Context-Aware Uncertainty Resolver}
\newcommand{\reasoner}{Plan–Guided Reasoner}

\usetikzlibrary{positioning, calc, fit}

\newcolumntype{L}[1]{>{\raggedright\arraybackslash}p{#1}}
\newcolumntype{C}[1]{>{\centering\arraybackslash}p{#1}}

\definecolor{kpOrange}{HTML}{F28C28}
\definecolor{kpGray}{HTML}{6E6E6E}
\definecolor{kpRed}{HTML}{C00000}

\definecolor{hdrBlue}{HTML}{BFE3F2}
\definecolor{rowGreen}{HTML}{CFE8C1}
\definecolor{subPurple}{HTML}{D8D0E8}

\definecolor{cellWhite}{HTML}{FFFFFF}

\newcommand{\storageicon}{\raisebox{-0.15em}{\includegraphics[height=1.1em]{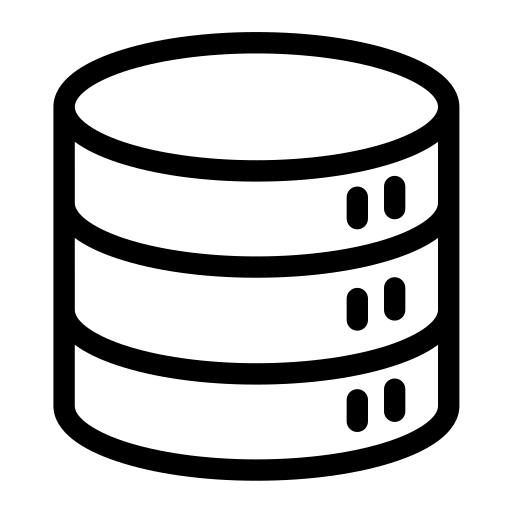}}}

\newcommand{\planicon}{\raisebox{-0.15em}{\includegraphics[height=1.1em]{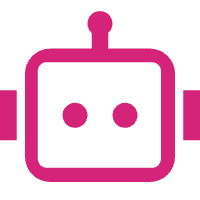}}}

\newcommand{\qaicon}{\raisebox{-0.15em}{\includegraphics[height=1.1em]{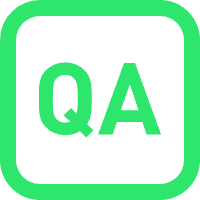}}}

\newcommand{\routericon}{\raisebox{-0.15em}{\includegraphics[height=1.1em]{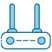}}}

\newcommand{\retrieveicon}{\raisebox{-0.15em}{\includegraphics[height=1.1em]{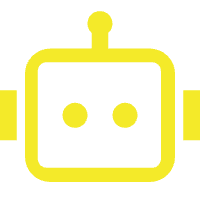}}}

\newcommand{\codingicon}{\raisebox{-0.15em}{\includegraphics[height=1.1em]{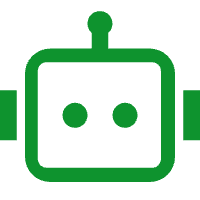}}}

\tcbset{
  mycompactbox/.style={
    enhanced,
    colback=white,
    colframe=gray!60,
    colbacktitle=gray!70,
    coltitle=white,
    fonttitle=\bfseries,
    boxrule=0.4pt,
    arc=0pt,
    left=3pt,right=3pt,top=2pt,bottom=2pt,
    boxsep=1pt,
    toptitle=1pt,bottomtitle=1pt,
    before skip=3pt,
    after skip=3pt,
  }
}

\begin{document}

\title{Cost-Effective Numerical QA over Semi-Structured Table: Structuring, Resolution, Planning
}

\author{
\IEEEauthorblockN{%
Feng Luo\textsuperscript{1,3}\textsuperscript{\textdagger},
Hui Luo\textsuperscript{2},
Zhifeng Bao\textsuperscript{3},
J.\,Shane Culpepper\textsuperscript{3},
Xiaoli Wang\textsuperscript{4},
Shazia Sadiq\textsuperscript{3}%
\thanks{\textsuperscript{\textdagger}This work was done while visiting The University of Queensland.}
}
\IEEEauthorblockA{%
\textsuperscript{1}RMIT University \quad
\textsuperscript{2}University of Wollongong \quad
\textsuperscript{3}The University of Queensland \quad
\textsuperscript{4}Xiamen University\\
feng.luo@student.rmit.edu.au,\;
huil@uow.edu.au,\;
zhifeng.bao@uq.edu.au,\;
s.culpepper@uq.edu.au,\\
xlwang@xmu.edu.cn,\;
shazia@eecs.uq.edu.au%
}
}

\maketitle

\begin{abstract}
Semi-structured tables encode rich semantic information through diverse layout elements, such as hierarchical row and column headers. 
Answering numerical questions over such data is challenging because it requires a joint effort of accurate structural understanding of tables, question uncertainty resolution, and complex query intents interpretation. 
Existing methods are rarely effective in handling the above multifaceted challenges in a holistic manner; moreover, they rely on LLMs without considering financial implications.

We propose \ourmethod{}, a cost-effective, LLM-powered framework featuring: 
(1) a Multiway Layered Table Structurer that converts a table into a multiway layered tree organized by structural elements;
(2) a Context-Aware Uncertainty Resolver that grounds underspecified phrases to context-coherent structural elements;
(3) a Plan–Guided Reasoner that decomposes complex query intents into a query plan for reasoning; 
(4) a Budget-Aware Bandit Router that learns per-instance LLM utilities and optimizes LLM selection under fixed monetary constraints. 
Experiments on three benchmarks show that \ourmethod{} improves both QA effectiveness and cost-effectiveness. Notably, the Table Structurer improves answer accuracy by 46\% over alternative tree construction methods, and Router achieves a 5\% average relative gain in answer accuracy compared with existing routing methods across budgets.



\end{abstract}


\section{Introduction} \label{sec_intro}
\begin{figure*}[t]
    \centering
    \includegraphics[width=0.85\linewidth]{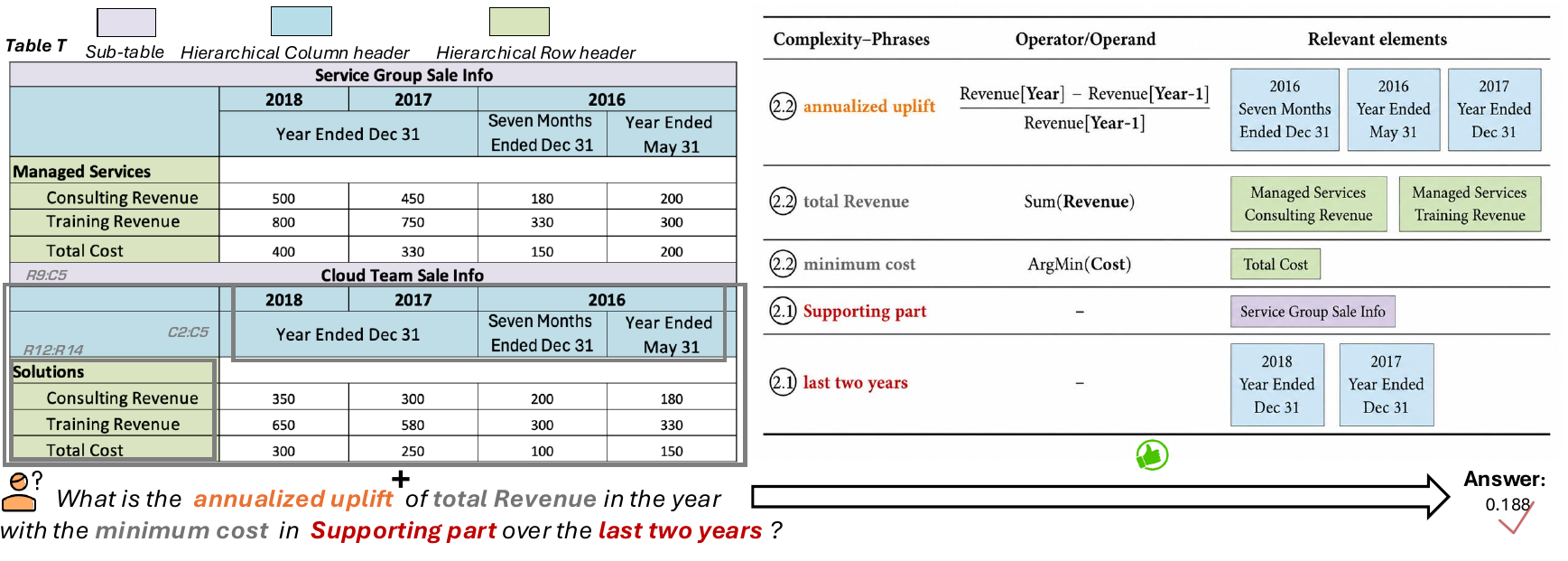}
 \captionsetup{skip=0pt}
    \caption{A running example illustrating our numerical semi-structured table QA problem.}
    \label{intro_example}
\vspace{-3pt}
\end{figure*}

Semi-structured tables flexibly capture complex layouts common in real-world data (e.g., financial reports \cite{srivastava2024evaluating} and electronic health records \cite{edin2023automated}).
Unlike flat tables, semantics can be (implicitly) encoded through rich layout elements, including the use of multiple subtables, hierarchical row headers, and hierarchical column headers \cite{wu2025realhitbench}. 
Notably, these tables often contain a substantial portion of numerical data (nearly 70\%\cite{zhao2022multihiertt}) and users frequently need to extract numerical insights from them, 
motivating the need for natural language interfaces to perform numerical question answering (QA) over such data.



In this paper, we study \emph{Numerical Semi-Structured Table QA}, which aims to accurately compute numeric answer for the question over semi-structured tables. Its complexity stems from two aspects: \emph{table side} and \emph{question side}.
On the table side, complexity mainly arises from  \emph{layout elements} that specify structural semantics for  interpreting values. Specifically, a single table may contain multiple subtables where a cell’s meaning is only recoverable by inferring its unique Hierarchical Header Sequence:
\textcircled{\tiny 1.1} \textit{hierarchical column headers}, where a column is specified using a nested column header sequence;
\textcircled{\tiny 1.2} \textit{hierarchical row headers}, where a row is specified using a nested row header sequence.
We illustrate this using the table $T$ shown in Fig.~\ref{intro_example}:



\begin{tcolorbox}[
  colback=gray!12,
  colframe=gray!45,
  boxrule=0.4pt,
  arc=2pt,
  left=4pt,right=4pt,top=2pt,bottom=2pt,
  boxsep=0.5pt,
  before skip=2pt,
  after skip=2pt
]
\small
\setlength{\parskip}{1pt}
\setlength{\parindent}{0pt}
\setlength{\baselineskip}{9pt}
$T$ contains two subtables: \emph{Service Group Sale Info} and \emph{Cloud Team Sale Info}.
Within \emph{Service Group Sale Info}:

\textcircled{\tiny 1.1} Determining the year in \emph{Year Ended May 31} requires inferring the column header sequence $(\emph{2016},\allowbreak\ \emph{Year Ended May 31})$.

\textcircled{\tiny 1.2} Determining the \emph{Total Cost} requires inferring the row header sequence $(\emph{Managed Services}, \emph{Total Cost})$.
\end{tcolorbox}

On the question side, complexity arises from:
\textcircled{\tiny 2.1}\ \textit{question uncertainty}, where questions are underspecified and require context-coherent grounding; 
\textcircled{\tiny 2.2}\ \textit{query intent}, where answering a question requires multi-step value grounding and reasoning, involving \emph{operands} and \emph{operators}. 
As shown in Fig.~\ref{intro_example}:


\begin{tcolorbox}[
  colback=gray!12,colframe=gray!45,
  boxrule=0.4pt,arc=2pt,
  left=4pt,right=4pt,top=2pt,bottom=2pt,
  boxsep=0.5pt,before skip=2pt,after skip=4pt
]
\small
\setlength{\parskip}{0pt}
\setlength{\parindent}{0pt}

\textcircled{\tiny 2.1}: ``last two years'' is underspecified and must be grounded to specific years, i.e., 2018 and 2017.\\[-1pt]
\textcircled{\tiny 2.2}: The query intent in ``annualized uplift of total revenue in the year with the minimum cost'' involves five steps: (1) locate data for the last two years, (2) rank these years by cost to find the minimum-cost year, (3) retrieve the values needed to instantiate the two operands for the uplift—the total revenue of that year and of the previous year, (4) aggregate the retrieved values, and (5) compute the uplift to produce the final answer. 
\end{tcolorbox}





Taken together, effectively addressing numerical semi-structured table QA requires a joint effort of 
\textbf{(C1)} \textit{accurate table structure understanding}, \textbf{(C2)} \textit{question uncertainty resolution}, and \textbf{(C3)} \textit{complex query intent interpretation}.
Existing approaches typically fall into three families: \emph{prompt-based} methods \cite{zhang2024tablellama} that employ pretrained LLMs to directly generate answers, \emph{code-driven} methods that translate questions into executable programs (e.g., SQL or Python), and \emph{representation-driven} methods that convert tables into representations (e.g., trees or graphs) and then use LLM-guided retrieval and reasoning over the representations for answering.
However, all three are rarely effective in handling the above multifaceted challenges in a holistic manner (see \S \ref{related_work_semi_qa}), as they simplify the table layout, or lack robust mechanisms for uncertainty resolution or complex query intent support.
For instance, the recent ST-Raptor~\cite{tang2025st} builds a hierarchical tree and uses LLMs to decompose a question into sub-questions and answer them over the tree.
It remains limited because:
(1) its tree omits key layout semantics (i.e., hierarchical row headers);
(2) it does not explicitly resolve underspecified phrases, causing decomposition errors to propagate to answering; and
(3) it answers sub-questions independently, ignoring dependencies that often arise in complex query intents.

Beyond effectiveness, existing methods rely on high-capability LLMs, overlooking the monetary cost of large-scale deployment.
A naive fix is to consider cheaper LLMs and route each instance (i.e., question and table pair) to an appropriate model to maximize accuracy under a fixed budget. A typical LLM routing paradigm is \emph{predict-then-optimize}  \cite{vsakota2024fly,mohammadshahi2024routoo,mei2025omnirouter}, which learns per-instance answer quality and cost predictions for candidate LLMs using the training set (e.g., thousands of instances), and then solves a constraint optimization problem to derive the routing policy. 
However, it is impractical for semi-structured table QA,
where limited training set (e.g., tens of instances) \cite{chai2025doctopus} makes predictions unreliable \cite{zeighami2025cut}. 
The prediction errors then propagate to optimization, yielding routing policies that neither satisfy the budget nor achieve good accuracy.
Consequently, we must address 
\textbf{C4:} \textit{How to select an LLM for each instance, aiming to maximize accuracy under a fixed budget, given only limited training data?}

\begin{figure}[t]
    \centering
    \includegraphics[width=0.9\columnwidth]{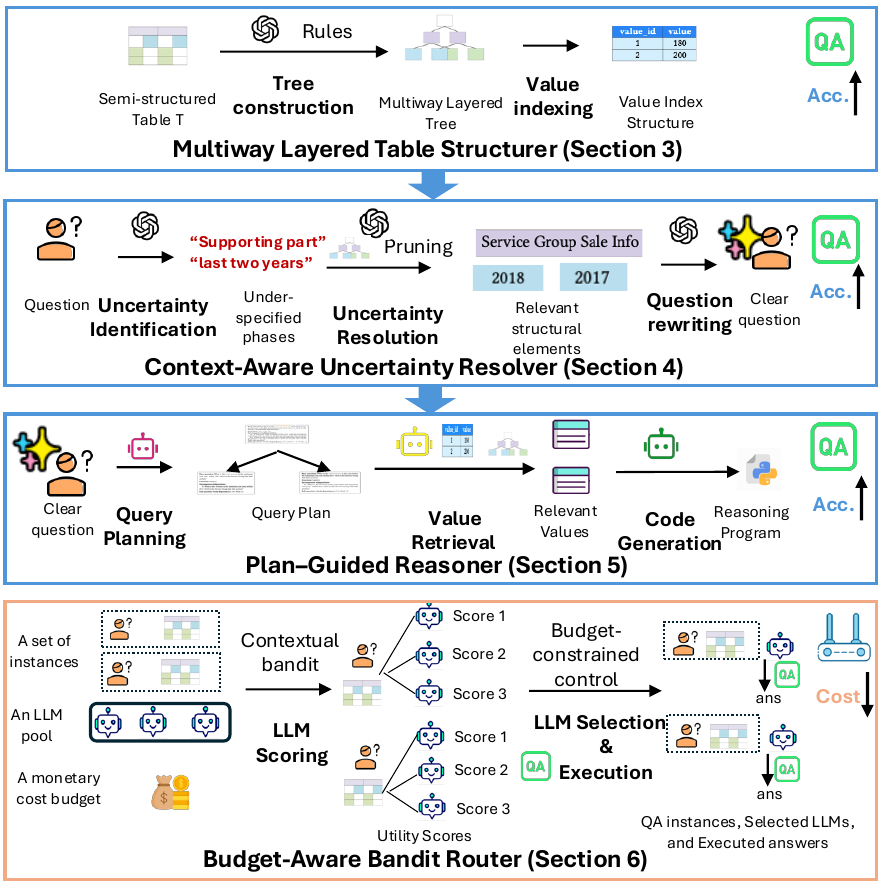}
    \captionsetup{skip=0pt}
    \caption{Overview of \ourmethod{}, containing a traceable QA pipeline \qaicon{}, and a cost-effective router \routericon{}.}
    \label{solution_overview}
    \vspace{-3pt}
\end{figure}

\smallskip
\noindent \textbf{Our Contributions.}
We introduce \ourmethod{}, a Cost-Effective, LLM-powered framework designed for Numerical Semi-Structured Table QA. \ourmethod{} integrates a traceable pipeline with a routing mechanism, as shown in Fig. \ref{solution_overview}. 

\begin{table*}[t]
\centering
\small
\begin{threeparttable}
\setlength{\tabcolsep}{5pt}
\begin{tabular}{@{}c l cccc cc@{}}
\toprule
\multirow{2}{*}{\textbf{Type}} &
\multirow{2}{*}{\textbf{Method}} &
\multicolumn{4}{c}{\textbf{Accurate table structure understanding}} &
\multirow{2}{*}{\makecell{\textbf{Question}\\\textbf{uncertainty}\\\textbf{resolution}}} &
\multirow{2}{*}{\makecell{\textbf{Complex}\\\textbf{query intent}\\\textbf{interpretation}}} \\
\cmidrule(lr){3-6}
& & \makecell{Multiple\\subtables} &
\makecell{Hierarchical\\column headers} &
\makecell{Hierarchical\\row headers} &
\makecell{Capability \tnote{2}} & & \\
\midrule
\multirow{1}{*}{Prompt-based} \tnote{1}
& TableLlama \cite{zhang2024tablellama}     & \xmark & \cmark & \cmark & $\star$     & $\star$     & $\star$ \\
\midrule
& TABFORMER \cite{dong2025reasoning}        & \xmark & \cmark & \cmark & $\star$     & $\star$     & $\star$ \\
\multirow{1}{*}{Code-driven}
& E5 \cite{zhang2024e5}                     & \xmark & \cmark & \cmark & $\star\star$ & $\star\star$ & $\star\star$ \\
\midrule
\multirow{3}{*}{Representation-driven}
& ST-Raptor \cite{tang2025st}               & \cmark & \cmark & \xmark & $\star$ & $\star$      & $\star$ \\
& TreeThinker \cite{wu2025realhitbench}     & \xmark & \cmark & \cmark & $\star\star$ & $\star\star$      & $\star\star$ \\
& GraphOTTER \cite{li2025graphotter}        & \xmark & \cmark & \cmark & $\star\star$ & $\star\star$ & $\star\star$ \\ 
\midrule
Hybrid \tnote{3} & \ourmethod{}        & \cmark & \cmark & \cmark & $\star\star\star$ & $\star\star\star$ & $\star\star\star$\\
\bottomrule
\end{tabular}
\begin{tablenotes}
      \footnotesize
      \item[1] We omit general-purpose foundation LLMs, as they are not tailored to semi-structured table QA, while the comparison results can be found in \S \ref{sec:exp}.
      \item[2] More $\star$ indicates stronger capability, based on the performance on our evaluation subsets requiring the corresponding capability (\S \ref{exp_qa_accuracy_overall} - \ref{exp_qa_robustness}).
      \item[3] We call it hybrid, as it contains structure modeling (as in representation-driven methods) and program generation (as in code-driven methods).

\end{tablenotes}
\captionsetup{skip=0pt}

\caption{Comparison of semi-structured table QA methods based on the required capabilities.}
\label{tab:research-gap}
\end{threeparttable}
\vspace{-3pt}
\end{table*}

\noindent$\bullet$~\textbf{Multiway Layered Table Structurer (\S \ref{method_structurer}).} To address \textbf{C1}, we propose a method that converts the table into a \emph{Multiway Layered Tree} organized by structural elements, enabling more accurate table structure understanding.

\noindent$\bullet$~\textbf{Context-Aware Uncertainty Resolver (\S \ref{method_resolver}).} 
  To address \textbf{C2}, we frame phrase grounding as an Extended Steiner Tree Problem, and use \emph{context-aware pruning} to jointly select a minimal, context-coherent set of structural elements to ground underspecified phrases and rewrite the question.
  
\noindent$\bullet$~\textbf{Plan–Guided Reasoner (\S \ref{method_reasoner}).}
   To address \textbf{C3}, we introduce the \emph{Operand-based} and \emph{Operator-based} decomposition strategies, to translate a complex query intent into a \emph{query plan} organized by a set of interdependent subintents.
   Guided by this plan, it grounds each subintent to the required values and composes an executable reasoning program, enabling robust support for complex intents.
   
\noindent$\bullet$~\textbf{Budget-Aware Bandit Router (\S \ref{method_router})}.
  To address \textbf{C4}, we reformulate the budget-constrained routing objective into an \emph{unconstrained Lagrangian surrogate objective}.
  Building on it, we propose a two-stage LLM routing strategy that (i) uses a contextual bandit algorithm to learn per-instance \emph{utility scores} for candidate LLMs, and (ii) enforces the global budget constraint by selecting the LLM and adaptively updating the cost penalty during execution, which makes the routing effective even with limited training data.

\noindent$\bullet$~\textbf{Experimental Evaluation (\S \ref{sec:exp}).} 
  Experiments on three benchmarks show that SemiBonsai improves both QA effectiveness and cost-effectiveness. Notably, the Table Structurer improves answer accuracy by 46\% over alternative tree construction methods, and the Router achieves a 5\% average relative gain in answer accuracy over existing routing approaches across monetary constraints.

\section{Problem Formulation and Literature Review} \label{problem_defn}
\subsection{Problem Definition}

As commonly found in real-world data (e.g., financial reports), a semi-structured table $T$ can be viewed as a 2D grid of cells arranged by rows and columns.
A \emph{column header} is a cell that labels a column, typically in the top rows.
A \emph{row header} is a cell that labels a row, typically in the left-most column.
Layout cues such as indentation and merged cells are often used to indicate that one header is nested under another.
We say two headers $h_1$ and $h_2$ satisfy a \emph{nesting relation} $h_1 \triangleright h_2$ if the layout indicates that $h_2$ is nested under $h_1$.
For example, in Fig.~\ref{intro_example}, the column header \emph{Year Ended May 31} is nested under the merged header \emph{2016}, i.e., $\emph{2016} \triangleright \emph{Year Ended May 31}$.

\noindent\textbf{Hierarchical headers.}
A \emph{hierarchical row header} is a maximal nested row header sequence
$p^r=(h^r_{1},\ldots,h^r_{L_{1}})$ of depth $L_1$ such that $h^r_i \triangleright h^r_{i+1}$ for all $i$, which identifies a row.
A \emph{hierarchical column header} is a maximal nested column header sequence
$p^c=(h^c_{1},\ldots,h^c_{L_{2}})$ of depth $L_2$ such that $h^c_i \triangleright h^c_{i+1}$ for all $i$, which identifies a column.
If no nesting relations exist, we treat its depth as $1$.
As shown in Fig. \ref{intro_example}, $(\textit{Managed Services},\ \textit{Consulting Revenue})$ is a hierarchical row header that identifies the 5th row, and
$(2018,\ \textit{Year Ended Dec 31})$ is a hierarchical column header that identifies the 2nd column.

\noindent\textbf{Subtables.}
A \emph{subtable} $s$ can be viewed as a rectangular region in $T$ delimited by a \emph{vertically contiguous} group of hierarchical row headers $\mathcal{P}^r_s$ and a \emph{horizontally contiguous} group of hierarchical column headers $\mathcal{P}^c_s$.
It may further have an optional merged \emph{title cell} whose text defines the subtable title $\mathcal{M}^t_s$, typically spanning all columns in $\mathcal{P}^c_s$ or all rows in $\mathcal{P}^r_s$.
We define the subtable metadata as $\mathcal{M}_s=(\mathcal{M}^t_s,\mathcal{P}^r_s,\mathcal{P}^c_s)$. The content of $s$ is denoted as $\mathcal{X}_s$, which is a set of cell values within the region delimited by $\mathcal{P}^r_s$ and $\mathcal{P}^c_s$.

For example, the grey boxes \texttt{C2:C5} and \texttt{R12:R14} in Fig.~\ref{intro_example} specify contiguous groups of hierarchical column headers and row headers, respectively, which together delimit the region \texttt{R9:C5} for a subtable titled \emph{Cloud Team Sale Info}.


\noindent\textbf{Semi-structured table.}
Accordingly, a semi-structured table is a collection of subtables within the 2D grid.
Formally, we represent it as
$T=\big(\mathcal{S},\{\mathcal{M}_s,\mathcal{X}_s\}_{s\in\mathcal{S}}\big)$,
where $\mathcal{S}$ is the set of subtables (($|\mathcal{S}|\ge 1$)).
We refer to all subtable metadata as its structural elements:
$\mathcal{E}(T)=\bigcup_{s\in\mathcal{S}}\mathcal{M}_s$.

For example, $T$ in Fig.~\ref{intro_example} contains two subtables, distinguished by titles \emph{Service Group Sale Info} and \emph{Cloud Team Sale Info}.
In this work, we assume that structural elements can be reliably indicated by layout cues; we do not consider cases where layout cues are misleading or relations can only be inferred using semantics across cell texts~\cite{cao2026orthogonal}. We focus on the common case where each subtable forms a rectangular region \cite{wu2025realhitbench}, and do not consider nested or interleaved subtables. 

\smallskip
\begin{defn}[Numerical Semi-Structured Table QA] \label{def_qa}
Given a semi-structured table $T$ and a numerical question $q$ (whose answer is a numeric value, obtained via arithmetic computation or aggregation),
it aims to compute the correct answer, $ans$, by learning the function $f: (q, T) \rightarrow ans$.
\end{defn}

\subsection{Literature Review}\label{related_work}
\subsubsection{Semi-structured table QA}\label{related_work_semi_qa}
Table question answering aims to answer a natural language question using a table and has been studied extensively~\cite{lu2025large}.
However, most prior work focuses on structured tables \cite{jiang2025tabdsr,poddubnyi2025query,xiao2025cents,zhang2024reactable,wang2024chain,zhang2025aixelask,jiang2026accurate,shahbazi2026omnitqa} and does not fully account for semi-structured tables due to their complex layouts \cite{zhang2025same}.
To bridge this gap, existing work for semi-structured table QA can be grouped into three categories (see Table~\ref{tab:research-gap}).

\noindent\textbf{Prompt-based methods}, such as TableLlama \cite{zhang2024tablellama} and GPT-5 \cite{gpt5}, use instruction prompting \cite{brown2020language} with foundation LLMs \cite{gpt5} or pretrained tabular LLMs  \cite{zhang2024tablellama} to generate the answer.
Since both table structure and question understanding are handled implicitly by the LLM itself, these methods remain limited in answering questions over the given table \cite{schwartz2024numerologic}.

\noindent\textbf{Code-driven methods} translate questions into executable programs such as Python or SQL, but semi-structured tables lack the explicit relational metadata needed for program generation.
Existing methods mainly address this through:
(1) \emph{table normalization}, which converts semi-structured tables into relational tables before program generation~\cite{nahid2024normtab,najpande2026quiett,dong2025relationalcoder,dong2025reasoning}.
For example, TABFORMER~\cite{dong2025reasoning} introduces a chain-of-transformation paradigm to normalize a semi-structured table and then generates SQL over the normalized table.
However, it discards key structural elements (e.g., multiple subtables), leading to structural information loss that propagates to SQL generation.
(2) \emph{structure summarization}, which summarizes the structure and generates programs based on the summaries.
For example, E5~\cite{zhang2024e5} leverages LLMs to produce summaries that capture hierarchy information and then generate Python programs.
However, such summaries often fail to capture subtable-level information (\S\ref{exp_ablation_study}) and it relies on LLMs to directly generate reasoning programs, without explicitly modeling uncertainty or complex intents.

\noindent\textbf{Representation-driven methods}, such as ST-Raptor~\cite{tang2025st}, TreeThinker~\cite{wu2025realhitbench}, and GraphOTTER~\cite{li2025graphotter}, first convert the table into a representation (e.g., tree/graph) and then retrieve relevant values and reasoning over them guided by LLMs.
ST-Raptor~\cite{tang2025st} constructs a hierarchical orthogonal tree using Visual Language Models (VLMs), then decomposes the question into sub-questions and answers them separately.
TreeThinker~\cite{wu2025realhitbench} uses an LLM to infer hierarchical row/column header trees and then prompts LLMs to retrieve values and reason over them.
GraphOTTER~\cite{li2025graphotter} models the table as an undirected graph linking neighboring headers/cells and answers by iteratively retrieving and expanding nodes before reasoning.
However, these methods remain limited: (1) their representations fail to capture all the structural elements (e.g., ST-Raptor omits hierarchical row headers), leading to incomplete structure understanding; (2)  they rely on LLM to handle uncertainty and intents implicitly, lacking explicit mechanisms for uncertainty resolution and intent understanding.

\subsubsection{LLM Routing}
Similar to our objective in the Router, existing LLM routing studies—which route each input query to one LLM—can be broadly categorized into three types based on their optimization objectives.
\textit{Quality-maximization routing} selects the model expected to yield the best answer quality for each query \cite{hu2024routerbench,chen2024routerdc,yan2026breaking}. 
\textit{Quality-constrained routing} aims to minimize the expected cost while ensuring the expected answer quality meets a required target over a query set~\cite{mei2025omnirouter,ding2025icml}.
\textit{Budget-constrained routing} maximizes expected answer quality while keeping the total cost within a given budget, which aligns with our cost-effective objective~\cite{panda2025adaptive,vsakota2024fly,wu2025port}.

Budget-constrained routing methods follow a predict-then-optimize paradigm \cite{vsakota2024fly,wu2025port}: they first learn predictors of per-instance answer quality and cost for each candidate LLM, then solve a constraint optimization problem to derive a routing policy. 
However, this paradigm is not applicable, as it assumes unavailable large training data. 
Moreover, because both cost and accuracy can vary across instances, it becomes even harder to learn accurate per-instance predictions. As a result, prediction errors can propagate into the optimization and degrade routing decisions.
PILOT~\cite{panda2025adaptive} can be a special case that trains a router online without requiring the training set, while it assumes online feedback of answer quality to update the routing decisions, which is unavailable in our setting.

\section{Multiway Layered Table Structurer} \label{method_structurer}

\begin{figure*}[t]
  \centering
  \includegraphics[width=0.9\linewidth]{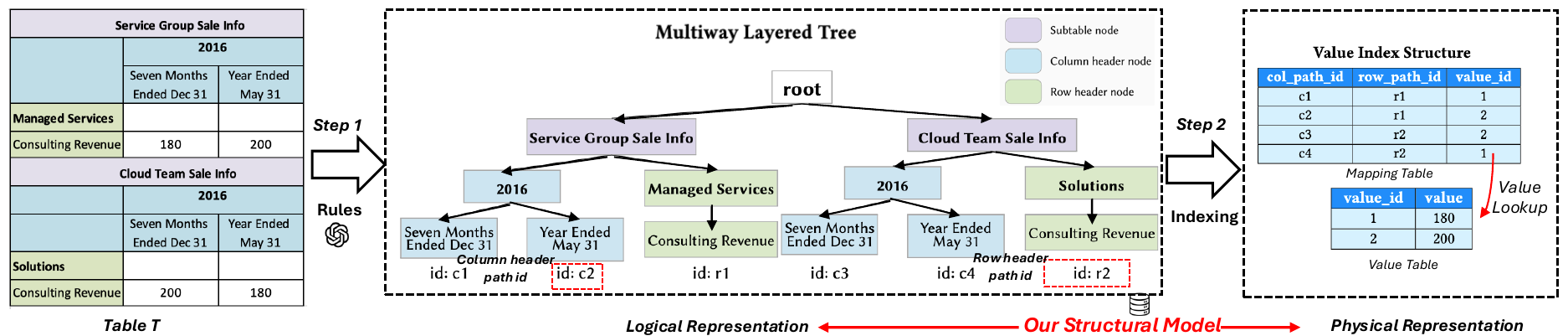}
  \captionsetup{skip=0pt}
  \caption{\small{\tablestructurer{}, which converts the table into a multiway layered tree and a storage-efficient value index structure.}}
  \label{fig:our_resp}
  \vspace{-3pt}
\end{figure*}

\noindent\textbf{Key issues.} Existing methods interpret table structural semantics via one of the following ways.
~\textbf{W1:} Tree construction -- building the hierarchy tree using an LLM/VLM \cite{tang2025st,wu2025realhitbench}.~\textbf{W2:} Graph modeling -- connecting neighboring headers and cells according to adjacency \cite{li2025graphotter}.~\textbf{W3:} Structure summarization -- prompting an LLM to output a structural summary \cite{zhang2024e5}.
\update{
~\textbf{W4:} Normalization -- applying the inferred structure transformation operations to obtain a relational table
\cite{nahid2024normtab,najpande2026quiett,dong2025relationalcoder,dong2025reasoning}.
}

However, these approaches exhibit two failure modes  
\ifcameraready
\else
(Appendix \S \ref{appendix_structural_error_patterns}):
\fi
\textit{missing structural information} (e.g., omitting header hierarchy) or \textit{spurious structural inference} (e.g., hallucinating incorrect hierarchy). 
\update{
These errors arise from limited tree, graph, and relational table designs \textit{that cannot fully capture diverse structural elements} and from LLMs’ \textit{unreliable identification for each type of element}.}
Moreover, we observe that existing tree/graph representations allocate a node per cell without considering empty cells or duplicated values, thereby \textit{incurring unnecessary storage overhead} for tables with high sparsity and redundancy \cite{bodensohn2025pvldb,luo2026vldb}.

\noindent\textbf{A novel structural model.}
To address them, we introduce a new structural model (Fig.~\ref{fig:our_resp}) consisting of: (1) a logical \emph{Multiway Layered Tree}, which
captures diverse structural elements in the tree structure; (2) a physical \emph{Value Index Structure}, which enables storage-efficient value lookup (\S \ref{structure_model_defn}).
Given table $T$, we first construct the layered tree with a \textit{rule-guided procedure} that exploits layout and styling cues to reliably capture diverse structural elements (\S\ref{schema_tree_construction}).
We then use the tree to index table content by de-duplicating cell values and materializing the Value Index Structure  (\S\ref{index_structure_construction}).

\subsection{A Logical--Physical Structural Model}\label{structure_model_defn}
\update{\noindent\textbf{Design rationale.}
As described in \S\ref{sec_intro}, semi-structured tables encode structural semantics through subtables, hierarchical row and column headers.
This motivates a \textit{logical multiway layered tree}, whose branches capture multiple subtables and header paths, and whose layers distinguish subtable, column-header, and row-header semantics, better preserving semantics than relational/graph representations.
Since storing cell values directly in the tree would introduce redundancy, we  store cell values separately in a \textit{physical index} that links each distinct value to its row and column header paths.}

\begin{cdefn}[Multiway Layered Tree]\label{def:schema_tree}
Given a semi-structured table
$T=\big(\mathcal{S},\{\mathcal{M}_s,\mathcal{X}_s\}_{s\in\mathcal{S}}\big)$,
it is a rooted, directed, multiway layered tree $G_T=(V,E)$ that organizes subtables and hierarchical headers.

\noindent\textbf{Nodes.}
$V$ contains a \textsf{root} node and three types of nodes:
(i) a \textsf{subtable} node for each $s\in\mathcal{S}$;
(ii) a \textsf{column header} node for each header  appearing in any hierarchical column header sequence in $\mathcal{P}^c_s$; and
(iii) a \textsf{row header} node for each header appearing in any hierarchical row header sequence in $\mathcal{P}^r_s$.

\noindent\textbf{Edges.}
$E$ encodes parent--child relations layer by layer:
(i) \textsf{root}$\rightarrow$\textsf{subtable} edges;
(ii) \textsf{subtable}$\rightarrow$\textsf{column header} \emph{and} \textsf{subtable}$\rightarrow$\textsf{row header} edges that connect each subtable to the top-level headers of its hierarchical column and row headers; and
(iii) \textsf{column header}$\rightarrow$\textsf{column header} \emph{and} \textsf{row header}$\rightarrow$\textsf{row header} edges that connect consecutive headers within each hierarchical header sequence.

A path starting at the \textsf{subtable} and ending at a leaf \textsf{column header}/\textsf{row header} node uniquely identifies a hierarchical column/row header in a subtable.
A row/column header subtree is a set of such paths sharing the same top-level row/column header node.
\end{cdefn}

\begin{ex}
As shown in Fig.~~\ref{fig:our_resp}, $\textsf{id{:}c2}$ identifies a path that specifies a hierarchical column header of subtable \emph{Service Group Sale Info}.
$\textsf{id{:}c1}$ and $\textsf{id{:}c2}$ belong to the same subtree.
\end{ex}

\smallskip
\begin{cdefn}[Value Index Structure]\label{def:value_index}
It consists of two relational tables: a \textsf{Value Table} and a \textsf{Mapping Table}.

\textsf{Value Table} has two columns,  $\textsf{value\_id}$ and $\textsf{value}$, where $\textsf{value\_id}$ is the primary key and $\textsf{value}$ stores each distinct raw cell value once.
\textsf{Mapping Table} has three columns $\textsf{col\_path\_id}$, $\textsf{row\_path\_id}$, and $\textsf{value\_id}$, where $\textsf{col\_path\_id}$ and $\textsf{row\_path\_id}$ identify root-to-leaf paths in $G_T$, and $\textsf{value\_id}$ refers to the corresponding cell value.
Its primary key is the composite key $(\textsf{col\_path\_id},\textsf{row\_path\_id})$, and its $\textsf{value\_id}$ is a foreign key referencing $\textsf{Value Table.value\_id}$.
\end{cdefn}

\subsection{Constructing the Layered Tree}\label{schema_tree_construction}
To construct a layered tree $G_T$, it requires identifying (i) the three types of nodes $V$, including subtable titles and headers; (ii) the parent--child edges $E$ that capture \textit{nesting relation} within hierarchical headers and the \textit{containment relation} between each subtable and its headers.
However, semi-structured tables do not specify such nodes and relations, and often express semantics via irregular layout cues (e.g., merged cells) and styling cues (e.g., boldface), which make reliable inference for semantics non-trivial.
To address this, we introduce a \textit{rule-guided two-stage procedure} to improve reliability: 
(1) use a VLM to identify candidate nodes, leveraging styling cues as additional signals, and (2) apply layout-driven rules to infer relations among candidates and construct a layered tree.

\noindent\textbf{Candidate node identification.}
Since the subtable titles and headers can be distinguished by styling cues beyond plain text, we leverage a visual rendering of $T$ and use a VLM to propose candidate nodes.
Specifically, we convert $T$ into HTML format, render it with a browser, and capture a high-resolution screenshot that preserves styling cues.
We then prompt VLM to propose candidate \emph{subtable title} and \emph{column/row header} from each cell text, forming an initial pool $V$.

\noindent\textbf{Rule-guided tree construction.} 
With the candidates, we can further prompt a VLM to infer nesting and containment relations from the rendered table image.
However, this can be unreliable for large tables (e.g., $>20$ rows) and complex layouts (e.g., $>10\%$ merged cells).
Because the number of candidates and possible relations grows quickly, making global relation inference error-prone and costly.
Therefore, we construct $G_T$ with a rule-guided procedure that scans the table and infers relations from layout cues, yielding a reliable Multiway Layered Tree, using three key steps:

\noindent\emph{\underline{Step 1}: Identifying nesting relations and constructing header subtrees.}
Starting from the header candidates, we scan the table grid in a fixed order (top-to-bottom by rows and left-to-right by columns) and match candidates to concrete cells using $n$-gram similarity (above a threshold $\tau$) to mitigate VLM hallucinations \cite{tang2025st}. Matched cells are added to two confirmed pools for row and column headers, and are instantiated as $\textsf{row header}$ or $\textsf{column header}$ nodes accordingly.
Meanwhile, we infer nesting relations and build edges by comparing each newly added header with previously confirmed headers in the same pool.
Specifically, for row headers, when a new node $h$ is added, we add edge from $h_1$ to $h$ if (i) \textit{indent rule}: $h$ is indented and $h_1$ is the confirmed nearest preceding less-indented header, or (ii) \textit{row-span rule}: $h_1$ is a confirmed merged cell whose row span covers $h$.
For column headers, when a new node $h$ is added, we add edge from $h_1$ to $h$ if (iii) \textit{col-span rule}: $h_1$ is a confirmed merged cell whose column span covers $h$.
After linking the header nodes, we then build header subtrees by treating each top-level header (i.e., a node with no incoming edge) as a root and taking the set of all descendant header nodes reachable from it as the subtree.

\noindent\emph{\underline{Step 2}: Identifying header subtree groups and inferring cross-adjacency.}
To identify contiguous header groups that can delimit subtables, we record the rows/columns covered by each header subtree and merge subtrees into \emph{maximal contiguous groups}.
Specifically, for column header subtrees, we union those whose covered columns together form an inseparable horizontal interval; for row header subtrees, we union those whose covered rows have an inseparable vertical interval. 
Based on these header subtree groups, we identify their \emph{cross-adjacency} via a \textit{corner-adjacent rule}:
a row header subtree group and a column header subtree group are \emph{corner-adjacent} if the rightmost column covered by the row group is immediately left of the leftmost column covered by the column group, or the topmost row covered by the row group is immediately below the bottommost row covered by the column group.

\noindent\emph{\underline{Step 3}: Identifying subtables via containment relations.}
We scan the grid and match title candidates to merged cells, instantiating each match as a title cell.
For each title cell, we apply a \textit{span-alignment rule}: 
if a title cell spans exactly the columns or rows covered by a corner-adjacent pair of a column header subtree group and a row header subtree group, we create a $\textsf{subtable}$ node and connect it to the root headers of the two subtree groups.
Finally, we connect each \textsf{subtable} node  to the \textsf{root} node, forming a layered tree.

\subsection{Storage-efficient Value Indexing}
\label{index_structure_construction}

Given the constructed layered tree $G_T$, we materialize the Value Index Structure by enumerating cell values indexed by structural elements and populating the \textsf{Value Table} and \textsf{Mapping Table}.
Specifically, we traverse all structural elements via root-to-leaf column and row header paths of  $G_T$.
For each column and row header path pair that uniquely identifies a cell in the original table grid, we fetch the raw value of that cell and populate the two relational tables as follows.
We first de-duplicate values in the \textsf{Value Table}: for each retrieved value not presented, we 
insert a tuple $(\textsf{value\_id}, \textsf{value})$, otherwise we reuse the existing $\textsf{value\_id}$.
We then insert $(\textsf{col\_path\_id}, \textsf{row\_path\_id}, \textsf{value\_id})$ into \textsf{Mapping Table}, recording the corresponding path pair and cell value.

\section{Context-Aware Uncertainty Resolver} \label{method_resolver}
\begin{figure}[t]
    \centering
    \includegraphics[width=\columnwidth]{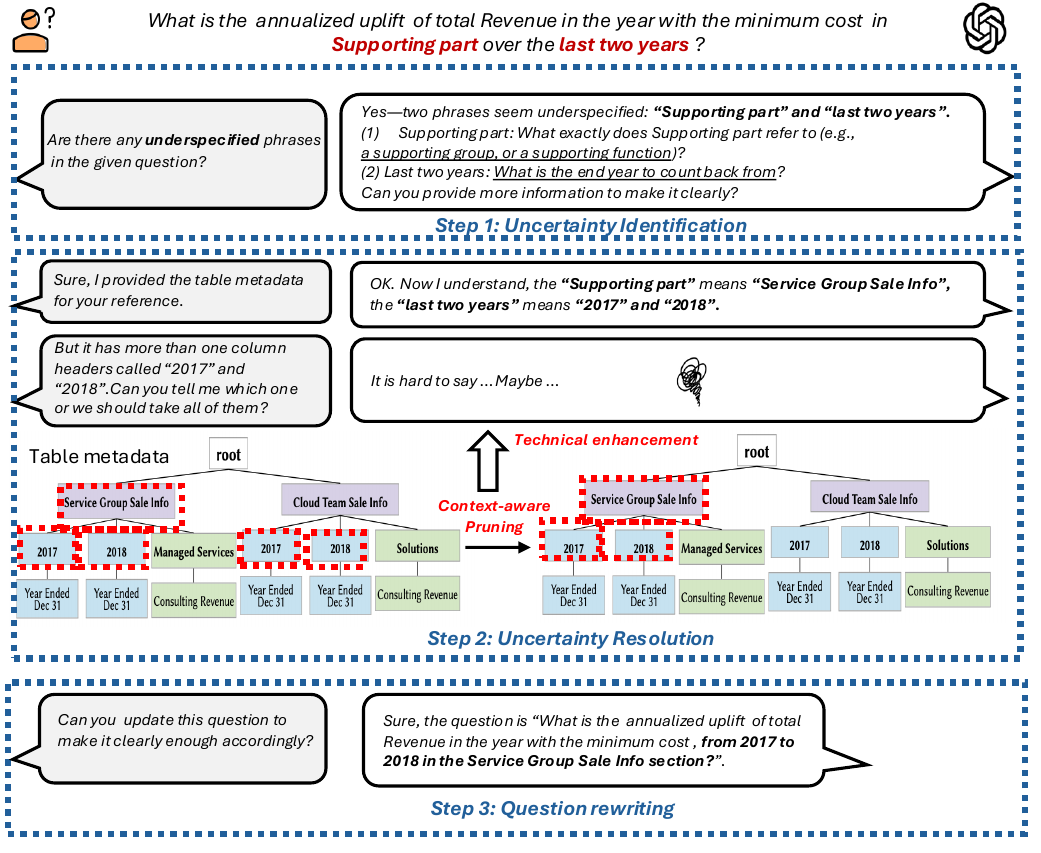}
    \captionsetup{skip=0pt}
    \caption{\uncertaintyresolver{}, which resolves the underspecified question into a fully specified one.}
    \label{method_resolution}
    \vspace{-3pt}
\end{figure}

\noindent\textbf{Key issues.} 
Existing methods either lack uncertainty resolution \cite{zhang2024tablellama,dong2025reasoning}, or handle it implicitly by independently mapping each phrase to structural elements
using LLMs \cite{zhang2024e5,wu2025realhitbench,li2025graphotter} or one-to-one embedding retrieval \cite{tang2025st}.
However, for questions with higher uncertainty (i.e., more under-specified phrases), these strategies often yield lower accuracy (\S\ref{exp_qa_robustness}) since they \textit{miss one-to-many mappings and cannot jointly ground multiple phrases under a coherent context.}
\begin{ex} \label{context_ex}
As shown in Fig.~\ref{method_resolution}, the question
contains two under-specified phrases: ``Supporting part'' and ``last two years''. Under a one-to-one assumption, it may miss that ``last two years'' corresponds to multiple headers (i.e., \emph{2018} and \emph{2017}). 
Moreover, resolving phrases independently can break the context coherence: ``last two years'' should be grounded within the subtable-level context \emph{Service Group Sale Info} identified by ``Supporting part'', while it retrieves \emph{2018} and \emph{2017} from an irrelevant subtable \emph{Cloud Team Sale Info}, introducing spurious elements and degrading accuracy.
\end{ex}

\noindent\textbf{A context-aware pruning strategy}. To address these issues, we propose a context-aware pruning strategy that ensures coherence by pruning irrelevant phrase groundings
, supporting one-to-many grounding (\S\ref{method_uncertainty_pruning}).
Building on it, \uncertaintyresolver{} converts an uncertain question into a fully specified one (Fig.~\ref{method_resolution}):
(1) uncertainty identification, (2) uncertainty resolution, and (3) question rewriting (\S \ref{method_uncertainty_query_rewriting}).

\subsection{Context-Aware Pruning}\label{method_uncertainty_pruning}
Intuitively, answering a question over a table requires identifying a set of structural elements that are not only relevant to its phrases, but also \emph{context-coherent} with each other under the table’s structure \cite{wang2020rat,kothyari2023crush4sql}, according to
(i) \emph{local context}, which constrains each phrase to structural elements structurally compatible  (e.g., within the same subtable) \cite{wang2020rat};
(ii) \emph{global context}, which prefers grounding all phrases using as few such local contexts as possible, yielding a more coherent grounding \cite{kothyari2023crush4sql}.
Accordingly, we propose \emph{context-aware pruning}, which aims to identify a minimal, context-coherent subset of structural elements that can jointly ground all phrases, thus pruning irrelevant candidates.

To achieve this, we model it as an extended Steiner Tree problem \cite{hwang1992steiner} over our layered tree:
\textit{given a candidate node set $\mathcal{C}(u)$ for each under-specified phrase $u$, we aim to find a minimum-size collection of connected components of $G_T$ such that, across the collection, every phrase is covered by at least one node.}
As it is computationally intractable in general, we adopt the 
faster approximation algorithm  \cite{mehlhorn1988faster} to obtain a solution.
Then, for each obtained connected component, we enumerate its root-to-leaf subpaths and lift each subpath to the corresponding longest subtable-to-leaf header path in $G_T$ that contains it.
We then use the union of structural elements on these lifted paths as the pruned results.

Using the strategy, we remove candidates outside selected components, as they are unsupported by local context (a connected component) and global context (a set of components).



\subsection{From an Uncertain Question to a Clear One}
\label{method_uncertainty_query_rewriting}

Given an input question $q$ with uncertainty, we need to resolve the uncertainty by converting it into
a fully specified question $\tilde{q}$ via explicitly grounding each under-specified phrase to relevant structural elements using $G_T$.

\noindent\textbf{Uncertainty identification.}
We call a phrase $u$ in $q$ \emph{under-specified} if it cannot be resolved without grounding to the table, e.g., it is referentially unclear (e.g., shorthand), or has an unclear spatial/temporal scope (e.g., ``last two years'') \cite{ding2025ambisql,zhang2025clear}.
Such uncertainty makes the question underspecified without a table.
We first prompt the LLM to extract under-specified phrases $\mathcal{U}=\{u_1,\ldots,u_{|\mathcal{U}|}\}$ from $q$ following \cite{ding2025ambisql}.

\noindent\textbf{Uncertainty resolution.}
Since under-specified phrases often require one-to-many grounding beyond embedding-based retrieval,
for each under-specified phrase $u_i$, 
we first leverage an LLM to identify a candidate set $\mathcal{C}(u_i)$ of nodes in $G_T$ that may ground $u_i$ (e.g., column header nodes ``2017'' and ``2018'' for ``last two years'') following \cite{zhang2025clear}.
To mitigate LLM hallucinations, we align each candidate to nodes in $G_T$ using $n$-gram similarity with threshold $\tau$, and discard unmatched candidates.
We then apply \emph{Context-Aware Pruning} to select relevant structural elements based on the aligned candidates.

For large tables where $G_T$ contains many nodes, 
which may reduce the effectiveness of LLM for candidate identification,
we support hybrid retrieval forming a small pool of nodes  \cite{zhang2025aixelask}.

\noindent\textbf{Question rewriting.}
Finally, we prompt an LLM to rewrite $q$ into $\tilde{q}$ by explicitly grounding each $u_i$ with the selected structural elements.
For example, in Fig.~\ref{method_resolution}, given the selected subtable \emph{Service Group Sale Info} and the hierarchical headers $(\emph{2017},\,\emph{Year Ended Dec 31})$ and $(\emph{2018},\,\emph{Year Ended Dec 31})$, the question is rewritten as:
``What is the annualized uplift of total Revenue in the year with the minimum cost, from 2017 to 2018 in the Service Group Sale Info section?''

\section{Plan–Guided Reasoner} \label{method_reasoner}
\noindent\textbf{Why do complex query intents break existing methods?}
A complex query intent generally involves multiple operators and operands.
However, prompt-based approaches often fail to infer all operands for computation~\cite{schwartz2024numerologic} and execute multiple operators~\cite{wang2024chain}.
Representation-driven methods are also brittle, as they either delegate the intent interpretation to the LLM or restrict it to a small set of predefined operators.
Code-driven methods are more promising because they compile the question into an executable reasoning program (e.g., SQL/Python) that can support operators and operands.
However, they synthesize the full program in one shot, which becomes fragile when multiple operands and operators are involved \cite{luo2026decomposition,yang2025code}.

\noindent\textbf{Our plan-guided reasoning framework.} To address this, we propose \reasoner{} (Fig.~ \ref{overview_reasoner}).
It first uses a planning agent to generate a \emph{Query Plan} by decomposing the complex query intent into subintents (\S\ref{method_reasoner_queryplanner}).
It then uses a grounding agent to ground each subintent with relevant values, and a coding agent to synthesize the executable reasoning program by following the subintent dependencies (\S \ref{method_reasoner_codegen}).
\subsection{Query Planning} \label{method_reasoner_queryplanner}
Answering numerical questions with complex intents often requires inferring multiple operands and composing multiple operators, making reasoning complicated.
To simplify it, we \emph{plan before solving}: we translate the question into a query plan as guidance. 
Specifically, we decompose the intent into a set of \emph{interdependent, minimal subintents}, each specified by required operands and operators and no longer decomposable.

\begin{figure}[t]
    \centering
    \includegraphics[width=\columnwidth]{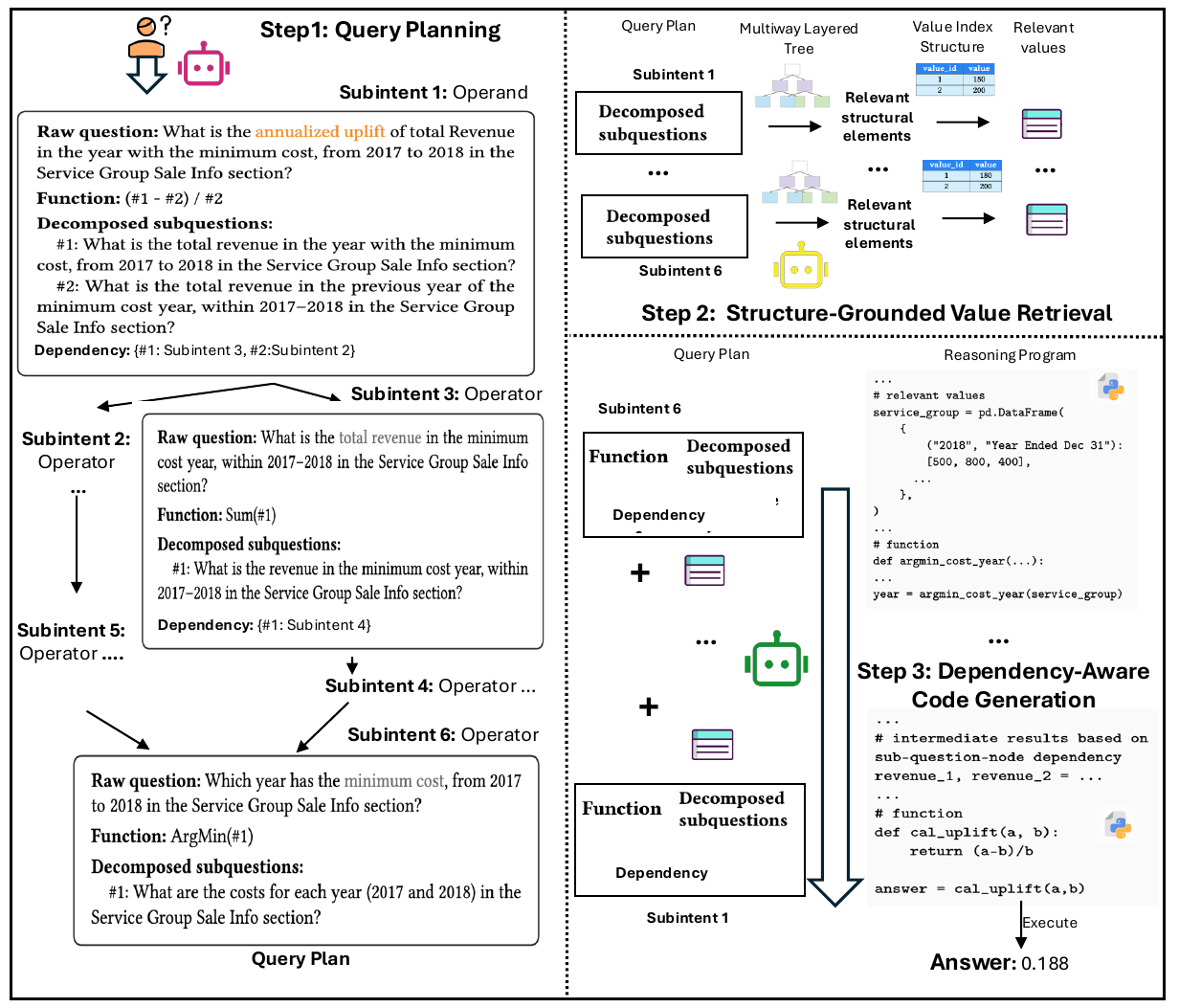}
    \captionsetup{skip=0pt}
    \caption{\reasoner{}, which synthesizes an executable reasoning program using the planning \planicon{}, grounding \retrieveicon{}, and coding \codingicon{} agents.}
    \label{overview_reasoner}
    \vspace{-3pt}
\end{figure}

 
\noindent\textbf{Subintent.}
A \emph{subintent} $\kappa =\langle \mathsf{func}, \mathsf{subqs}, \mathsf{rawq}, \mathsf{rel} \rangle$, specifying (1) which computation (i.e., a math expression) and operator (including aggregation, comparison, ranking, filtering, grouping) should be performed, using \texttt{Function} ($\mathsf{func}$),  (2) what table evidence is needed to supply its operands and parameters,  using \texttt{Decomposed subquestions} ($\mathsf{subqs}$), (3) \texttt{Raw question}, which is the natural language description of this subintent ($\mathsf{rawq}$), and (4) optional \texttt{Dependency}, which reveals the dependencies between subintents ($\mathsf{rel}$).
For example, in Fig.~~\ref{overview_reasoner}, Subintent~1 encodes the computation $(\#1-\#2)/\#2$ and lists subquestions to supply operands $\#1$ and $\#2$; the dependency indicates that these operands are provided by Subintent~2 and Subintent~3.

\noindent\textbf{Subintent-to-plan.}
Given the rewritten question $\tilde{q}$, we first initialize a root subintent $\kappa_0$ with $\tilde{q}$ as its $\mathsf{rawq}_0$, and then use an LLM-based planning agent to perform decomposition based on $\tilde{q}$ by choosing one of two ways:
(i) \emph{Operand-based decomposition.}
When $\tilde{q}$ involves a numerical computation, the agent first infers the computation as $\mathsf{func}_0$ and then decomposes $\tilde{q}$ into a set of subquestions $\mathsf{subqs}_0$, where each subquestion is designed to supply one distinct operand required by $\mathsf{func}_0$;
(ii) \emph{Operator-based decomposition.}
When $\tilde{q}$ involves a multi-operator chain, the agent extracts the first operator to apply as $\mathsf{func}_0$, and rewrites $\tilde{q}$ into $\mathsf{subqs}_0$ by splitting it into subquestions that produce the parameters for $\mathsf{func}_0$.

After obtaining $\kappa_0$, we push it into a stack $\mathsf{St}$ and iteratively perform decomposition until $\mathsf{St}$ is empty. Specifically,
at each step, we inspect the top Subintent $\kappa\in\mathsf{St}$.
We call $\kappa$ \emph{minimal} if none of its subquestions in $\mathsf{subqs}$ can be further decomposed by either operand-based or operator-based decomposition; in this case, we pop $\kappa$ and add it to an Subintent set $\mathcal{K}$.
Otherwise, for each decomposable subquestion in $\mathsf{subqs}$, we apply operand-based or operator-based decomposition to obtain a new Subintent $\kappa'$ similar to $\tilde{q}$.
We then update $\kappa$'s dependency $\mathsf{rel}$ to link the corresponding decomposable subquestion in $\kappa$ to $\kappa'$, and push $\kappa'$ into $\mathsf{St}$.
Once $\mathsf{St}$ is empty, we construct the query plan by organizing $\mathcal{K}$ using the dependencies.

Compared with existing planning approaches~\cite{LanWBLJLLHQH26,zhang2024reactable,zhang2025aixelask}, our query planning offers two advantages:  (1) It jointly specifies operands and operators, while prior work mainly performs operation-based decomposition~\cite{poddubnyi2025query,zhang2025aixelask}, which is insufficient 
\ifcameraready
\else
for numerical questions (Appendix~\S\ref{appendix_exp_llm_comparison}).
\fi
(2) It explicitly models dependencies among sub-intents, rather than producing unstructured sequential decompositions~\cite{wang2024chain,zhang2024reactable}, making it suitable for non-linear reasoning \cite{zhang2025aixelask}.

\subsection{From Plan to Execution} \label{method_reasoner_codegen}
With the produced query plan, we still need to derive the final answer by: (1) retrieving relevant values as evidence for $\mathsf{subqs}$ in each Subintent, and (2) executing the subintents in dependency order to compose intermediate results progressively.

\noindent\textbf{Structure-Grounded value retrieval}.
We traverse the query plan top-down and process each Subintent’s $\mathsf{subqs}$.
For any independent $\mathsf{subqs}$ (i.e., it is not linked to another Subintent via $\mathsf{rel}$), we retrieve its required table evidence as follows.
We first invoke an LLM-based grounding agent to extract key phrases and identify relevant nodes in the constructed layered tree for each phrase.
We then apply the context-aware pruning strategy (\S\ref{method_uncertainty_pruning}) to select the most relevant structural elements.
Finally, we use the selected elements’ path identifiers to query the Value Index Structure: we look up the corresponding tuples in the \textsf{Mapping Table} and join their \textsf{value\_id} to the \textsf{Value Table} to obtain the values.
We return these values together with the elements as the grounded evidence for the $\mathsf{subqs}$.

\noindent\textbf{Dependency-Aware code generation}.
After obtaining all the required evidence, we generate an executable reasoning program by following the query plan bottom-up.
We use Python (rather than SQL) for the reasoning program since our setting lacks an explicit database schema required for SQL generation.
Specifically, we invoke an LLM-based coding agent to synthesize a subprogram for each Subintent: it (i) implements the Subintent’s $\mathsf{func}$ and (ii) obtains $\mathsf{func}$’s operands/parameters either from retrieved evidence or from intermediate variables produced by dependent child subintents (as specified by $\mathsf{rel}$).
Finally, the agent composes these subprograms together following the dependency via $\mathsf{rel}$ and feeding each child’s output variable into the corresponding operand/parameter of its parent, forming a reasoning program.
If execution fails, we re-invoke coding agent with the error message to repair the code, up to a maximum of 3 attempts.
\section{Budget-Aware Bandit Router} \label{method_router}
Up to now, we have an LLM-powered QA pipeline that can produce an answer.
However, relying on high-cost LLMs makes the pipeline impractical under budget constraints.
Thereby, we incorporate smaller LLMs and study \emph{budget-aware LLM routing for Numerical Semi-Structured Table QA} (\S\ref{method_routing_problem_setup}),  which turns our pipeline into a cost-effective solution. 
We then introduce our two-stage routing strategy (\S\ref{method_routing_approach}).

\subsection{Problem Setup} \label{method_routing_problem_setup}
Let $\mathcal{D}=\{(q_k,T_k)\}_{k=1}^{N}$ be a set of instances, where each instance consists of a question $q_k$ and a semi-structured table $T_k$. 
Let $\mathcal{L}$ be a pool of candidate LLMs.
A routing policy is a mapping $\pi:\mathcal{D}\rightarrow\mathcal{L}$ that assigns one LLM to each instance.

\noindent\textbf{Cost and answer accuracy.}
Given an LLM \(\ell\in\mathcal{L}\), we instantiate our learned QA function \(f\) (Definition~\ref{def_qa}) with \(\ell\), denoted as
\(f_{\ell}:(q, T)\rightarrow \hat{ans}\).
Applying \(f_{\ell}\) on \((q, T)\) also incurs a monetary cost \(c({\ell}, q,T)\in\mathbb{R}_{\ge 0}\).
Answer accuracy is computed by comparing \(\hat{ans}\) with the ground-truth answer \(y\) using a numerical equivalence function \(\mathrm{Eq}(\cdot,\cdot)\), following~\cite{wang2024enhancing}:
$
a(\ell, q,T)=\mathbf{1}\!\left[\mathrm{Eq}\!\big(\hat{ans},\,y\big)\right]\in\{0,1\}.
$

\smallskip
\begin{defn}[Budget-Aware LLM Routing for Numerical Semi-Structured Table QA] \label{sec6_costeffective_obj}

Given query instances $\mathcal{D}$, an LLM candidate pool $\mathcal{L}$, and a total budget $B$, the goal is to learn a routing policy that selects one LLM per instance to maximize average answer accuracy under the budget constraint:
{\small
\begin{align}
\pi^\star \;=\; \arg\max_{\pi:\mathcal{D}\to\mathcal{L}} \;& \frac{1}{N}\sum_{k=1}^{N} a\!\left(\pi(q_k,T_k),\, q_k,\, T_k\right) \\
\text{s.t.}\;& \sum_{k=1}^{N} c\!\left(\pi(q_k,T_k),\, q_k,\, T_k\right)\ \le\ B,
\end{align}
}
where $\pi(q_k,T_k)$ is the assigned LLM for the instance $(q_k,T_k)$.
\end{defn}

\subsection{Two-Stage Routing: Contextual Bandit Scoring and Budget-Constrained Control}\label{method_routing_approach}
As discussed in \S\ref{sec_intro}, a limited training set makes the predict-then-optimize routing paradigm unreliable for our Budget-Aware LLM Routing problem
, because it is hard to obtain accurate per-instance predictions of both answer accuracy and cost from only a few examples.
A naive fix is to synthesize more training instances \cite{chai2025doctopus}. 
However, this is unlikely to work well for semi-structured table QA, as the complexity of answering each instance varies widely, making it hard to generate instances with similar complexity.

To address this, we apply a \emph{Lagrangian relaxation} \cite{fisher1981lagrangian} to our budget-constrained routing objective 
\ifcameraready
\else
(Appendix \S\ref{app:routing_theory})
\fi
: we replace the global budget constraint with a cost penalty~$\lambda$ and obtain an \emph{unconstrained Lagrangian surrogate objective} that scores each candidate LLM by the \textit{utility}
$u_\lambda(\ell;q,T)=a(\ell,q,T)-\lambda\,c(\ell,q,T)$.
For a fixed~$\lambda$, it then selects the LLM with the largest utility score $u_\lambda$, as a per-instance LLM selection rule induced by the relaxed objective.
This avoids learning precise per-instance predictions of accuracy and cost
; instead, we directly learn to rank candidate LLMs by their utility for each instance, which is more robust with limited training data \cite{dimakopoulou2021online}.
Since the Lagrangian relaxation replaces the hard budget constraint with a soft cost penalty, and may not enforce the budget, we therefore adapt~$\lambda$ to enforce the global budget constraint, building on these learned utilities.
Accordingly, our routing follows two stages: (1) learn utility-based scoring for candidate LLMs; (2) enforce the budget by adaptively updating~$\lambda$ during execution.

\noindent\textbf{LLM scoring.}
To effectively learn the instance-dependent utility scores, we use the contextual bandit algorithm \cite{beygelzimer2011contextual}, encoding each instance as a context vector $x$ by concatenating the question with a flattened list of layered tree nodes for $T$ and embedding the resulting text using a lightweight DistilBERT encoder \cite{sanh2019distilbert}.
Using the training set as offline bandit feedback---i.e., we only observe the utility of the LLM computed from each instance's true accuracy and cost---we train the contextual bandit to estimate the expected utility of each candidate LLM conditioned on $x$.
At inference time, for a new instance $(q,T)$, we compute $x$ and output a utility estimate $\widehat{u}_\lambda(\ell;x)$ for every $\ell\in\mathcal{L}$, then route to the LLM with the largest estimated utility, 
$\ell^*=\arg\max_{\ell\in\mathcal{L}} \widehat{u}_\lambda(\ell;x)$.

\noindent\textbf{Budget-Constrained control.}
To satisfy the global budget constraint $B$, given a sequence of instances $\{(q_k,T_k)\}_{k=1}^{N}$, we maintain the cumulative cost $C_{k-1}$ and a cost-penalty parameter $\lambda_k$ before routing the $k$-th instance.
For the current instance, we score each candidate $\ell\in\mathcal{L}$ as $\widehat{u}_{\lambda_k}(\ell;x_k)$ with the current penalty $\lambda_k$,
and select the LLM with the highest utility score.
After executing the chosen LLM and observing its cost \(c_k\), we update \(C_k=C_{k-1}+c_k\) and adjust the penalty via a dual update \(\lambda_{k+1}\leftarrow\big[\lambda_k+\eta\cdot(c_k-\tfrac{B}{N})\big]_+\), where \(\eta>0\) is a step size and \([\cdot]_+\) clamps negative values to zero.

\textbf{Rationale behind our framework.}
\ourmethod{} starts with the Table Structurer to interpret and cache the table’s structural semantics, making them reusable for questions over the same table.
It then applies the Uncertainty Resolver to ground underspecified phrases and produce a clear question, which is necessary for the Reasoner to reliably translate the complex intents into a query plan.
Finally, since they are LLM-powered, we place a Router on top of them for cost-effectiveness.
\section{Experimental Evaluation} \label{sec:exp}
\noindent\underline{\textit{QA effectiveness, robustness, and efficiency.}}
We assess the effectiveness, robustness, and efficiency of \ourmethod{}:

\begin{itemize}[leftmargin=*,noitemsep]
    \item Q1: How effective is \ourmethod{} overall?
    \item Q2: How robust is \ourmethod{} to varying table side complexity and question side complexity?
    \item Q3: How efficient is \ourmethod{} in runtime/token usage?
    \item Q4: How does each component affect effectiveness?
\end{itemize}

\noindent\underline{\textit{Cost-effectiveness.}}
We assess the cost-effectiveness:
\begin{itemize}[leftmargin=*,noitemsep]
    \item Q5: Can the \router{} in \ourmethod{} keep the total monetary cost within a given budget while maximizing the answer accuracy?
\end{itemize}

\subsection{Experimental Setup}
\subsubsection{Datasets}
We evaluate on three widely used semi-structured table QA benchmarks: HiTab \cite{cheng2022hitab}, MultiHiertt \cite{zhao2022multihiertt}, and RealHiTBench \cite{wu2025realhitbench}.
We choose them because their tables have much higher numerical value ratios, i.e., the fraction of cells with numerical values (0.72/0.68/0.79),
which provides more numerical questions that match our setting.
While other benchmarks are more text-dominated, with a much lower numerical value ratio.
For example, TempTabQA~\cite{gupta2023temptabqa} has 0.04, InfoTabs~\cite{gupta2020infotabs} has 0.03, WikiTQ~\cite{pasupat2015compositional} has 0.26, and SSTQA~\cite{tang2025st} has 0.40,
making them less suitable for evaluation.
\update{Note that tables in existing semi-structured table QA benchmarks are generally small (fewer than 100 rows) \cite{tang2025st}.}

Using these benchmarks, we derive numerical-only subsets by \update{filtering questions with numeric ground-truth answers over single table from HiTab and MultiHiertt validation splits  and RealHiTBench's numerical-reasoning split}.
We denote them as HiTab-Num, MultiHiertt-Num, and RealHiTBench-Num.

\noindent\textbf{Complexity of Questions and Tables.}
To assess the impact of question and table complexity to  QA effectiveness,  we curate three complex subsets using the above datasets: \(\qsubset{}\) (complex-question subset), \(\tsubset{}\) (complex-table subset), and \(\qtsubset{}\) (both complex question and complex table), by automatically classifying questions and tables into simple or complex.
Instead of using a heuristic complexity score with a fixed threshold, which can be sensitive to interactions among factors such as the number and nesting depth of structural elements,
\ifcameraready
\else
we train feature-based classifiers to learn a data-driven decision boundary (Appendix~\S\ref{appendix_complex_subset_construction}).
\fi

Specifically, we (i) construct table-side and question-side complexity features, (ii) build small labeled pools manually, and (iii) train two binary classifiers (for table-side and question-side complexity), which are then applied to all datasets.

Table~\ref{tab:dataset_stats} summarizes key factors that contribute to table-side and question-side complexity.
For tables, we measure (i) the number of subtables, (ii) the mean nesting depth of hierarchical column headers, and (iii) the mean nesting depth of hierarchical row headers. 
For questions, we measure (i) question uncertainty via the under-specified ratio (the fraction of key phrases not directly matched with annotated relevant structural elements), (ii) the number of grounded operands, and (iii) the number of required operators.

\begin{table}[t]
\centering
\small
\setlength{\tabcolsep}{3pt}
\renewcommand{\arraystretch}{1.15}
\definecolor{lightblue}{RGB}{170,205,240}

\resizebox{\columnwidth}{!}{%
\begin{tabular}{l r r r r r r r r}
\toprule
\multirow{2}{*}{Dataset} &
\multirow{2}{*}{\#T} &
\multicolumn{3}{c}{\textbf{Table}} &
\multirow{2}{*}{\#Q} &
\multicolumn{3}{c}{\textbf{Question}} \\
\cmidrule(lr){3-5}\cmidrule(lr){7-9}
& &
\#Sub &
Avg. ColDepth &
Avg. RowDepth &
&
Avg. UnderSpec &
\#Operands &
\#Operators \\
\midrule
HiTab-Num        & 80  &
\cellcolor{lightblue!20}1.00 &
\cellcolor{lightblue!72}2.20 &
\cellcolor{lightblue!35}1.51 &
203 &
\cellcolor{lightblue!51}0.51 &
\cellcolor{lightblue!20}2.07 &
\cellcolor{lightblue!28}3.26 \\
MultiHiertt-Num  & 158 &
\cellcolor{lightblue!20}1.00 &
\cellcolor{lightblue!72}1.61&
\cellcolor{lightblue!35}1.56&
240 &
\cellcolor{lightblue!74}0.66 &
\cellcolor{lightblue!64}2.61 &
\cellcolor{lightblue!59}3.56 \\
RealHiTBench-Num & 191 &
\cellcolor{lightblue!80}1.34 &
\cellcolor{lightblue!20}2.00 &
\cellcolor{lightblue!20}1.51&
276 &
\cellcolor{lightblue!20}0.31 &
\cellcolor{lightblue!56}2.52 &
\cellcolor{lightblue!20}3.18 \\
\midrule
\qsubset{}       & 201 &
\cellcolor{lightblue!39}1.11 &
\cellcolor{lightblue!23}1.89 &
\cellcolor{lightblue!34}1.50 &
295 &
\cellcolor{lightblue!80}\textbf{0.70} &
\cellcolor{lightblue!80}\textbf{2.81} &
\cellcolor{lightblue!80}\textbf{3.76} \\
\tsubset{}       & 130 &
\cellcolor{lightblue!45}\textbf{1.14} &
\cellcolor{lightblue!80}\textbf{2.25} &
\cellcolor{lightblue!80}\textbf{2.01} &
206 &
\cellcolor{lightblue!54}0.53 &
\cellcolor{lightblue!52}2.46 &
\cellcolor{lightblue!32}3.30 \\
\qtsubset{}      & 83  &
\cellcolor{lightblue!43}\textbf{1.13} &
\cellcolor{lightblue!67}\textbf{2.17} &
\cellcolor{lightblue!78}\textbf{1.99} &
107 &
\cellcolor{lightblue!80}\textbf{0.70} &
\cellcolor{lightblue!70}\textbf{2.69} &
\cellcolor{lightblue!46}\textbf{3.43} \\
\bottomrule
\end{tabular}%
}
\par
\vspace{2pt}
\scriptsize
\#Sub: \#subtables; ColDepth/RowDepth:nesting depth; UnderSpec:under-specified ratio.
\captionsetup{skip=0pt}
\caption{Dataset statistics.}
\label{tab:dataset_stats}
\vspace{-3pt}
\end{table}

\subsubsection{Baselines}
We compare \ourmethod{} with representative semi-structured table QA baselines from three categories.
Prompt-based methods using GPT-5 \cite{gpt5} and TableLlama \cite{zhang2024tablellama}.
Code-driven methods include E5 \cite{zhang2024e5} and TABFORMER \cite{dong2025reasoning}. Since there is no publicly available implementation for TABFORMER, we reimplement its pipeline by adopting NormTab \cite{nahid2024normtab} for table normalization and using an LLM-based SQL generator.
Representation-driven methods include ST-Raptor \cite{tang2025st}, TreeThinker \cite{wu2025realhitbench}, and GraphOTTER \cite{li2025graphotter}.
The details of each method can be found in \S\ref{related_work}.

\subsubsection{Evaluation metrics}
To measure answer accuracy on numerical questions, we adopt Arithmetic Exact Match (EM) following \cite{wang2024enhancing}, which measures the percentage of predictions whose numeric values match the ground truth. 
Since equivalent answers may appear in different formats (e.g., 10\% vs. 0.1), we apply an LLM-based value equivalence check for comparison following \cite{tang2025st}.
\update{
This LLM-based equivalence check matches human judgments in 95\% of cases (evaluated on 40 sampled prediction--gold answer pairs).}

\subsubsection{Implementation}
We use GPT-4.1 as the primary backend LLM for all baselines and \ourmethod{}, and as the VLM for table structurer. The matching threshold $\tau$ is set to $0.9$. 
Since methods support different input table formats, we provide each table in two equivalent formats (HTML and Excel): HTML for prompt-based and code-driven methods, and Excel for \ourmethod{} to align with representation-driven methods.
\ifcameraready
\else
The prompts used by \ourmethod{} are included in Appendix \S\ref{appendix_prompts}. 
\fi
Router implementation details of \ourmethod{} are provided in \S\ref{exp_cost_effectiveness}. 
Our source code is available at \cite{sourcecode}.

\captionsetup[table]{skip=2pt, belowskip=0pt}

\begin{table}[t]
\centering
\small
\setlength{\tabcolsep}{1pt}

\newcommand{\bestHitabf}{0.618}
\newcommand{\bestMultif}{0.588}
\newcommand{\bestRealf}{0.631}
\newcommand{\relimpf}[2]{\fpeval{round(((#1-#2)/#2)*100,0)}\%}

\begin{minipage}{\columnwidth}
\centering
\begin{tabular}{l c c c}
\toprule
Method & HiTab-Num & MultiHiertt-Num & RealHiTBench-Num\\
\midrule
TableLlama  & 0.333 & 0.096 & 0.079 \\
GPT-5       & 0.603 & 0.338 & 0.502 \\
\midrule
TABFORMER   & 0.123 & 0.225 & 0.220 \\
E5          & 0.422 & \underline{0.588} & 0.627 \\
\midrule
ST-Raptor    & 0.319 & 0.354 & 0.308 \\
TreeThinker  & 0.554 & 0.221 & 0.519 \\
GraphOTTER   & \underline{0.618} & 0.279 & \underline{0.631} \\
\midrule
\ourmethod{} & 
\textbf{0.724}$^{\scriptsize(+\relimpf{0.724}{\bestHitabf})}$ &
\textbf{0.708}$^{\scriptsize(+\relimpf{0.708}{\bestMultif})}$ & \textbf{0.744}$^{\scriptsize(+\relimpf{0.744}{\bestRealf})}$ \\
\bottomrule
\end{tabular}

\captionsetup{type=table}
\caption{Answer accuracy across three benchmarks using EM. 
Relative gains in parentheses compare \ourmethod{} with the strongest baseline (underlined).
}
\label{tab:overall_answer_accuracy}
\end{minipage}

\vspace{3pt}

\begin{minipage}{\columnwidth}
\centering

\newcommand{\bestHitab}{0.430}
\newcommand{\bestMulti}{0.514}
\newcommand{\bestReal}{0.481}

\newcommand{\relimp}[2]{\fpeval{round(((#1-#2)/#2)*100,0)}\%}

\begin{tabular}{l c c c}
\toprule
Method & \qsubset{} & \tsubset{} & \qtsubset{} \\
\midrule
TableLlama   & 0.095 & 0.131 & 0.075 \\
GPT-5        & 0.341 & 0.434 & 0.389 \\
\midrule
TABFORMER    & 0.121 & 0.127 & 0.093 \\
E5           & \underline{\bestHitab} & \underline{\bestMulti} & \underline{\bestReal} \\
\midrule
ST-Raptor    & 0.224 & 0.255 & 0.176 \\
TreeThinker  & 0.301 & 0.401 & 0.287 \\
GraphOTTER   & 0.388 & 0.467 & 0.389 \\
\midrule
\ourmethod{} &
\textbf{0.646}$^{\scriptsize(+\relimp{0.646}{\bestHitab})}$ &
\textbf{0.693}$^{\scriptsize(+\relimp{0.693}{\bestMulti})}$ &
\textbf{0.714}$^{\scriptsize(+\relimp{0.714}{\bestReal})}$ \\
\bottomrule
\end{tabular}

\captionsetup{type=table}
\caption{Answer accuracy using EM on the three subsets.}
\label{tab:complex_answer_accuracy}
\end{minipage}
\vspace{-3pt}
\end{table}

\subsection{QA Effectiveness (Q1)} \label{exp_qa_accuracy_overall}
We first report overall QA effectiveness on three datasets and then evaluate the complex subsets.

\subsubsection{Main results} 
Table \ref{tab:overall_answer_accuracy} summarizes answer accuracy on three benchmarks.  We observe that:
(1) \ourmethod{} significantly outperforms all baselines, achieving 18\% average improvement over the best baseline, highlighting the benefits of jointly modeling table structure, uncertainty, and complex intents.
(2) No single baseline is consistently strong, as they prioritize different capabilities. Representation-driven methods better capture structures (e.g., GraphOTTER on HiTab-Num), while code-driven methods better handle complex intents and uncertainty (e.g., E5 on MultiHiertt-Num).

\subsubsection{Results on complex questions and tables}
Table~\ref{tab:complex_answer_accuracy} reports accuracy on complex subsets.
We observe that:
(1) On \qsubset{}, \ourmethod{} yields a \(50\%\) improvement over the best baseline, suggesting that explicitly resolving question-side complexity is critical.
(2) On \tsubset{}, \ourmethod{} also achieves clear gains, showing that effective table-structure modeling is necessary for complex tables.
(3) Overall, \ourmethod{} consistently attains the highest accuracy across all subsets, indicating its robust performance.

\begin{figure}[t]
  \centering
  \includegraphics[width=0.7\columnwidth]{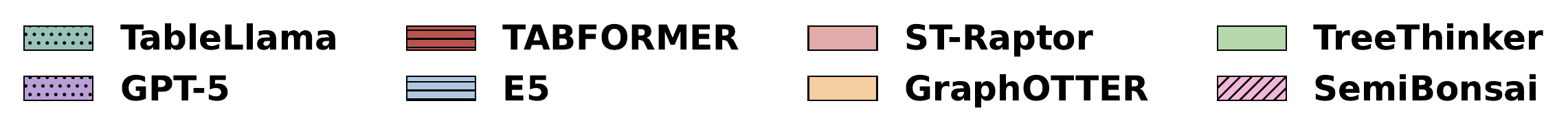}

  \begin{subfigure}[t]{0.32\linewidth}
    \centering
    \includegraphics[width=\linewidth,trim=9mm 10mm 0mm 8mm]{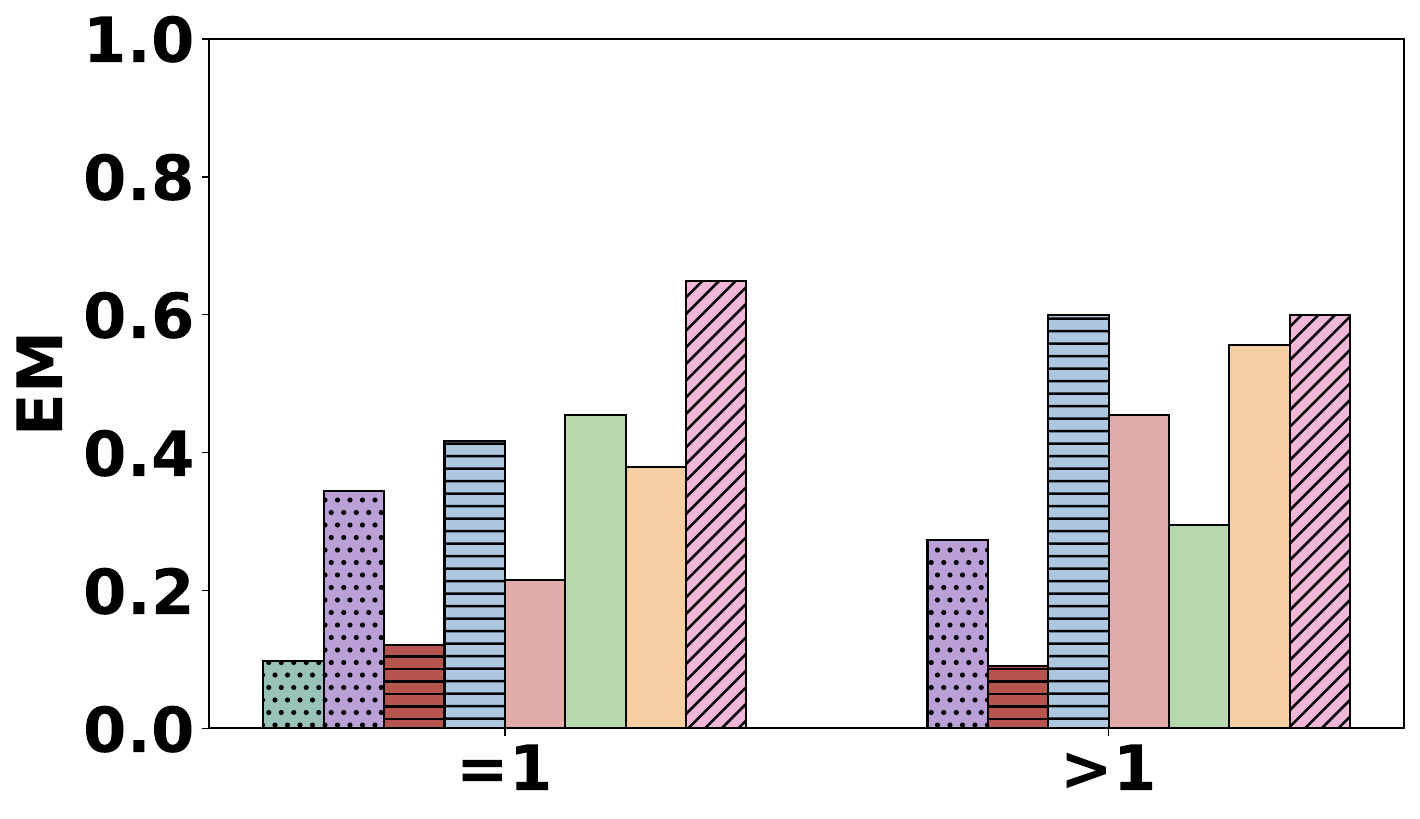}
    \caption{\scriptsize Subtables.}
    \label{fig:tbl_factor_subtable}
  \end{subfigure}
  \hfill
  \begin{subfigure}[t]{0.32\linewidth}
    \centering
    \includegraphics[width=\linewidth,trim=9mm 10mm 0mm 8mm]{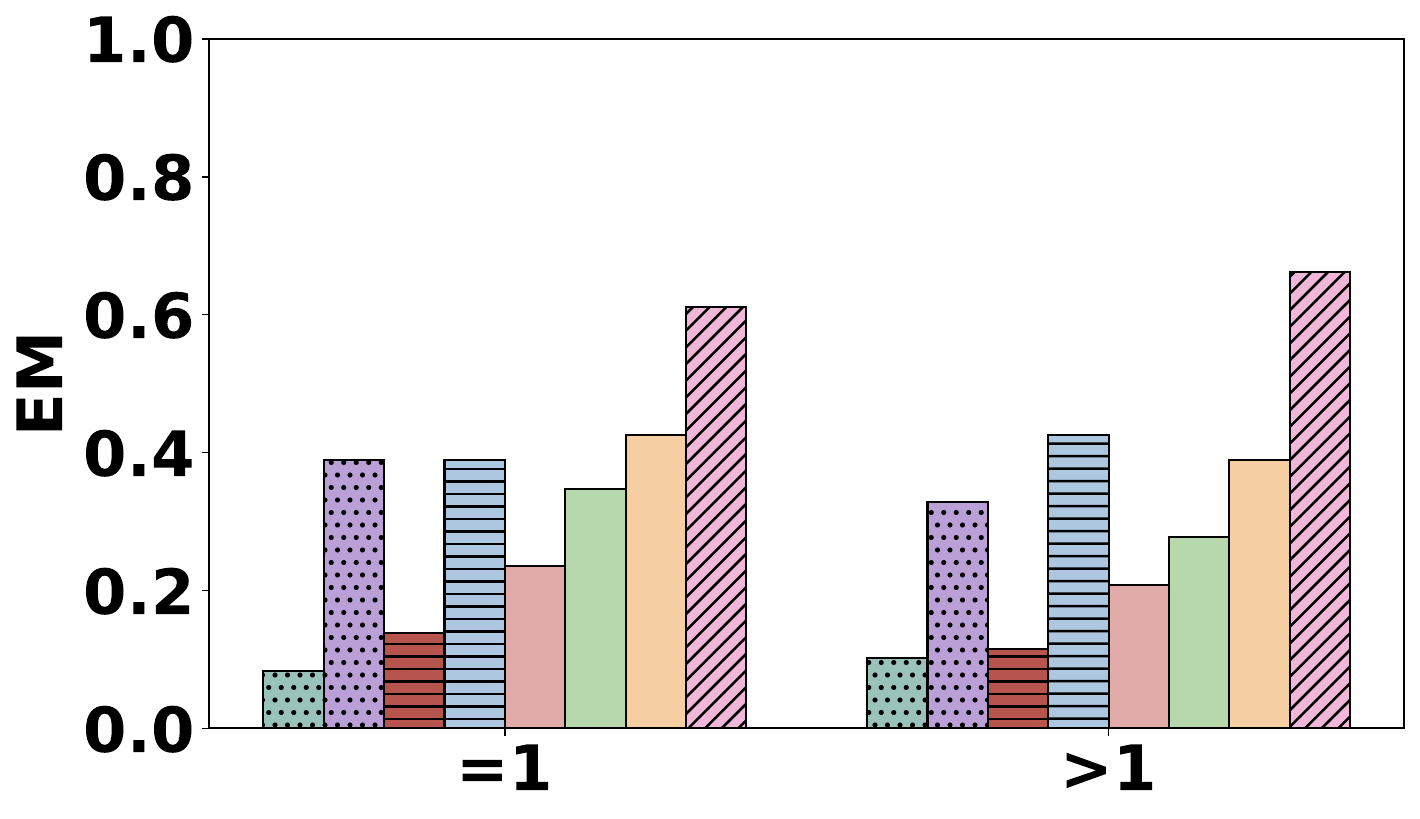}
    \caption{\scriptsize Column header depths.}
    \label{fig:tbl_factor_colheader}
  \end{subfigure}
  \hfill
  \begin{subfigure}[t]{0.32\linewidth}
    \centering
    \includegraphics[width=\linewidth,trim=9mm 10mm 0mm 8mm]{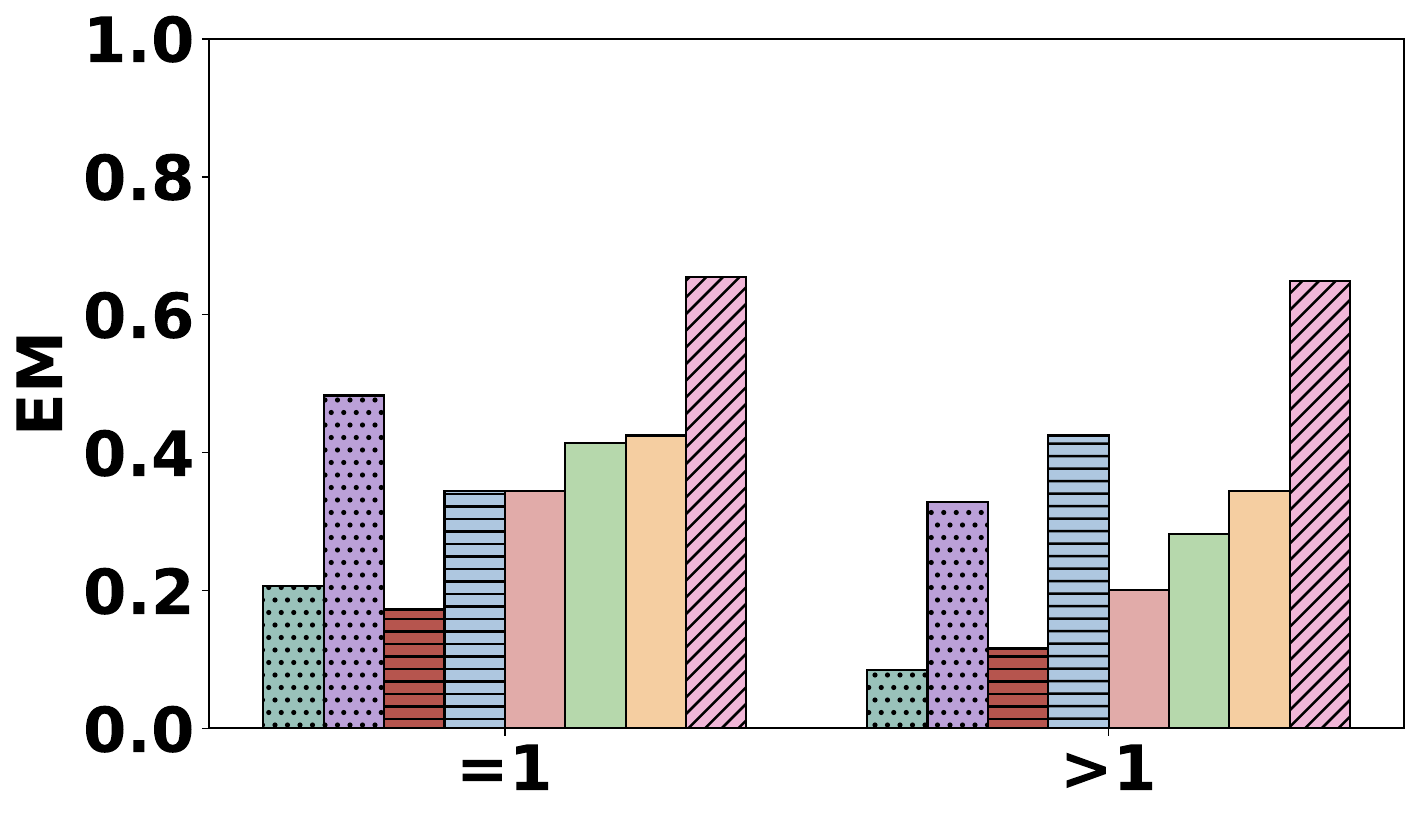}
    \caption{\scriptsize Row header depths.}
    \label{fig:tbl_factor_rowheader}
  \end{subfigure}
  \captionsetup{skip=0pt}
    \caption{Robustness under table-side factors on \(\qsubset{}\).} 
  \label{fig:table_side_complexity_impact}
  \vspace{-3mm}
\end{figure}
\begin{figure}[t]
  \centering

  \begin{subfigure}[t]{0.35\linewidth}
    \centering
    \includegraphics[width=\linewidth,trim=9mm 10mm 0mm 8mm]{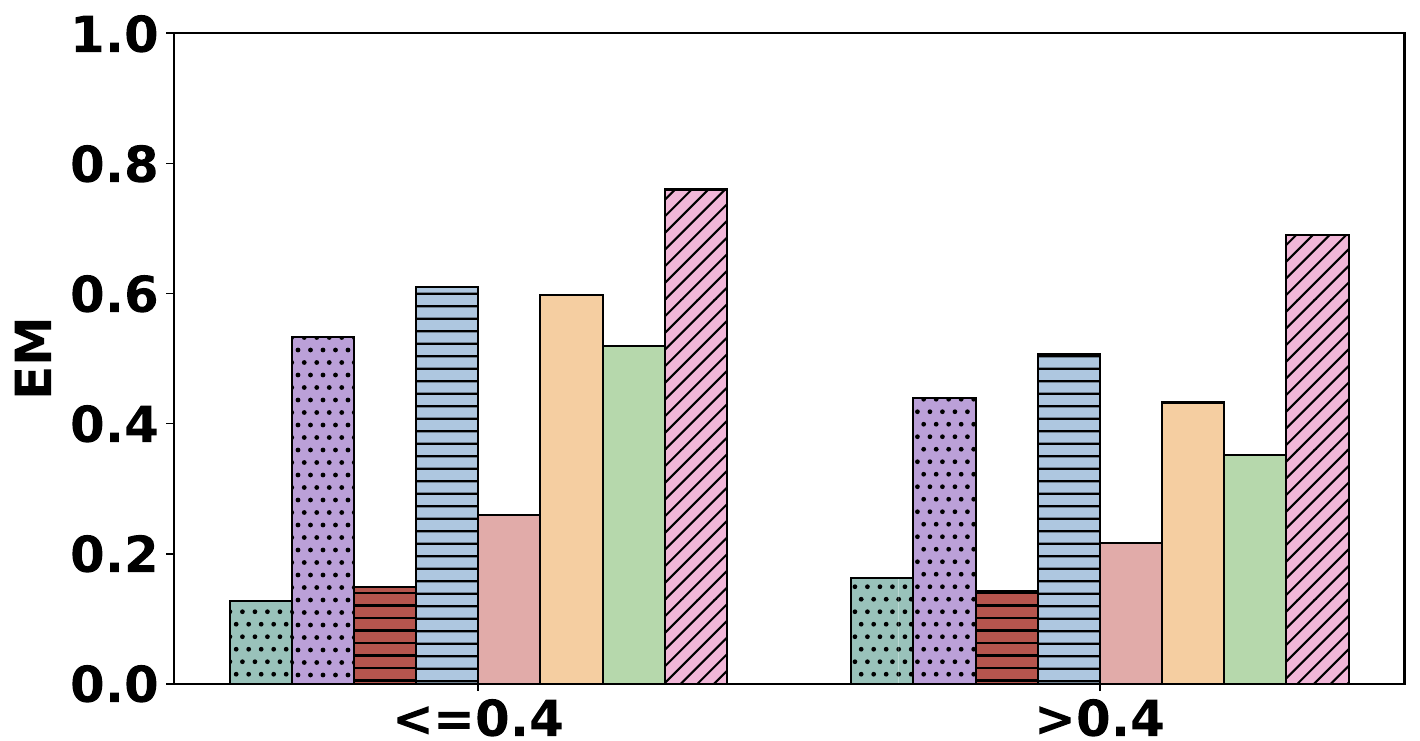}
    \caption{\scriptsize Question Uncertainty}
    \label{fig:qside_schema}
  \end{subfigure}
  \hspace{2mm}
  \begin{subfigure}[t]{0.35\linewidth}
    \centering
    \includegraphics[width=\linewidth,trim=9mm 10mm 0mm 8mm]{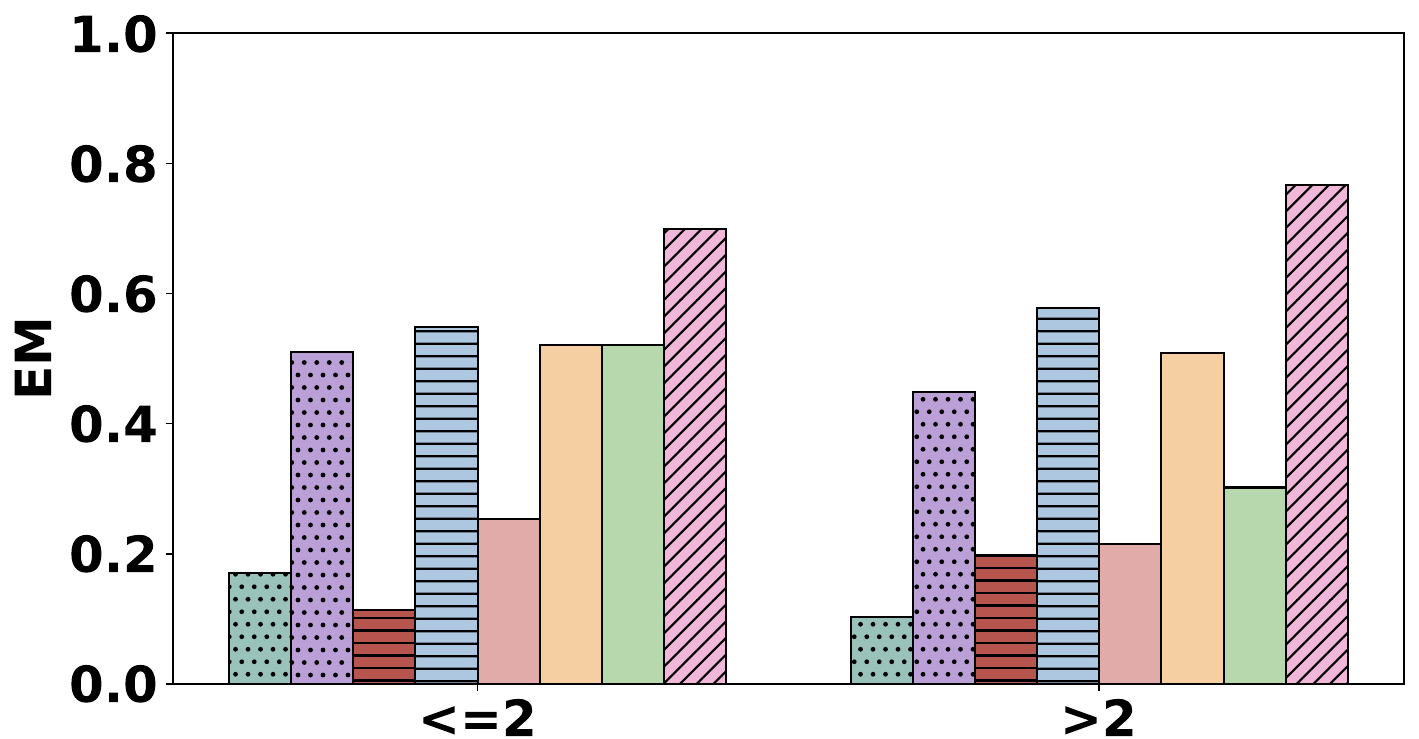}
    \caption{\scriptsize Number of operands}
    \label{fig:qside_num}
  \end{subfigure}

  \vspace{1mm}

  \begin{subfigure}[t]{0.35\linewidth}
    \centering
    \includegraphics[width=\linewidth,trim=9mm 10mm 0mm 8mm]{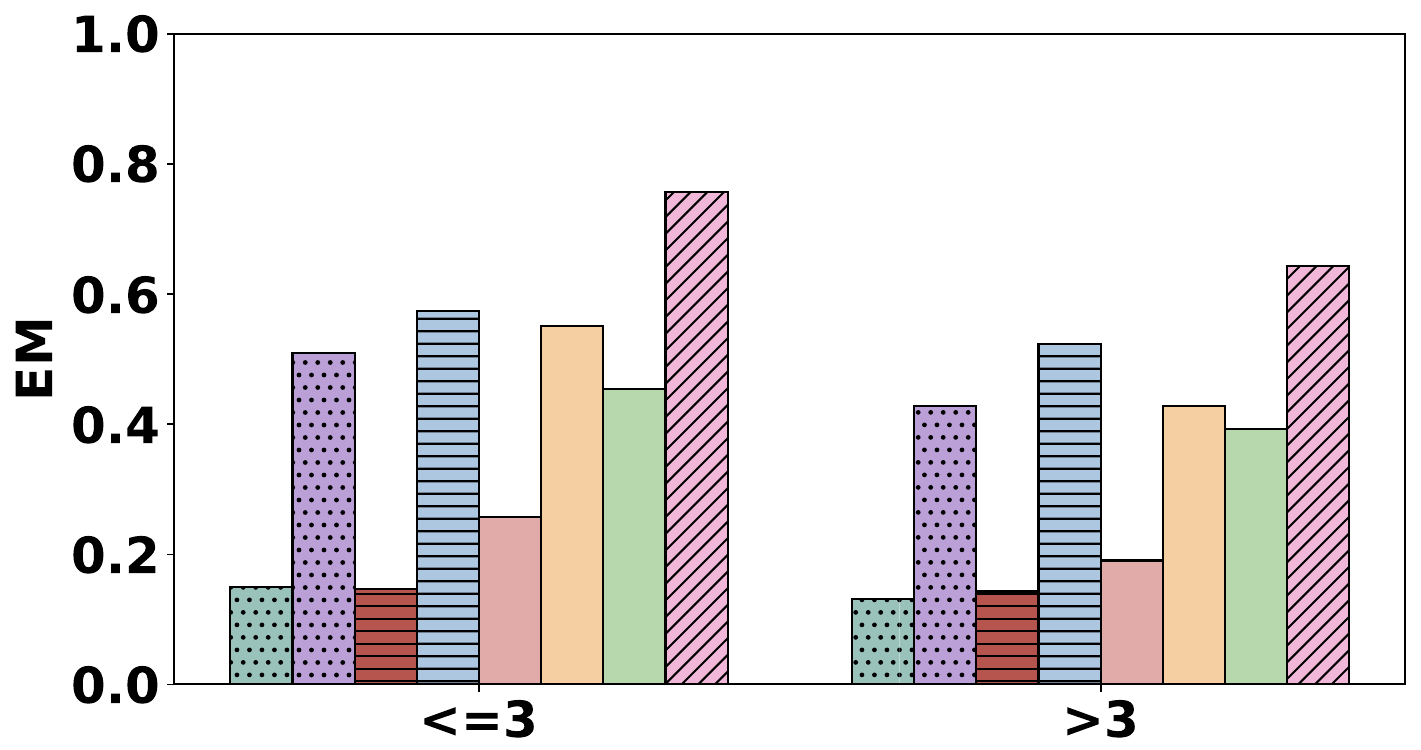}
    \caption{\scriptsize Number of operators}
    \label{fig:qside_reasoning}
  \end{subfigure}
  \hspace{2mm}
  \begin{subfigure}[t]{0.35\linewidth}
    \centering
    \includegraphics[width=\linewidth,trim=9mm 10mm 0mm 8mm]{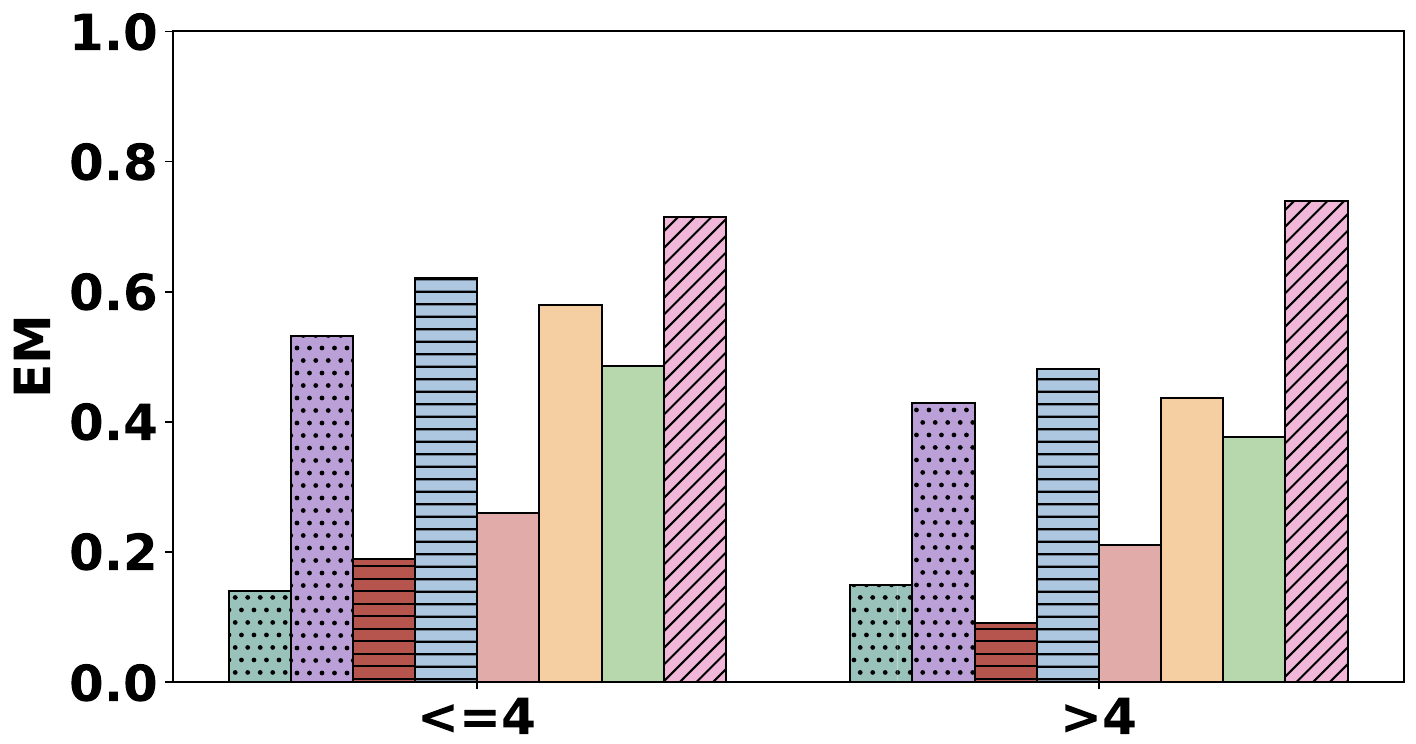}
    \caption{\scriptsize Number of structural elements}
    \label{fig:qside_elements}
  \end{subfigure}

  \captionsetup{skip=0pt}
  \caption{Robustness under question-side factors on \small{\tsubset{}}.}
  \label{fig:question_side_complexity_impact}
  \vspace{-3mm}
\end{figure}

\subsection{Robustness (Q2)} \label{exp_qa_robustness}
In this subsection, we analyze robustness with respect to fine-grained table-side and question-side factors that can increase complexity and affect QA effectiveness. To focus on these complex cases while controlling the other side,  we evaluate table-side factors on \(\qsubset{}\) and question-side factors on \(\tsubset{}\).
In addition to the factors in Table~\ref{tab:dataset_stats}, we also include question-side \emph{value grounding complexity}, approximated by the number of required structural elements, since uncertain and complex-intent questions often require grounding more elements to locate necessary values.

\subsubsection{Impact of table-side factors}
Since table-side factors are discrete, we use rule-based splits: single vs.\ multiple subtables, and hierarchical headers when the mean nesting depth exceeds \(1\).
Fig.~\ref{fig:table_side_complexity_impact} shows that (1) most baselines degrade noticeably as table-side complexity increases, because they cannot reliably capture diverse structural elements, which downgrade answer accuracy.
(2) \ourmethod{} is the most robust across all splits, because its Layered Tree offers a more reliable table structure, enabling more accurate value grounding even with multiple subtables and deeper header hierarchies.
As existing benchmarks contain few tables with multiple subtables, we further stress-test \ourmethod{} on \qsubset{} using augmented variants with 3, 5, or 7 subtables.
Its accuracy remains stable across these settings 
\ifcameraready
\else
(details are in Appendix~\S\ref{appendix_exp_llm_comparison}).
\fi

\subsubsection{Impact of question-side factors}
Since question-side factors are continuous and more dispersed, we split the subset into low vs.\ high groups using the median of its empirical distribution.
Fig.~\ref{fig:question_side_complexity_impact} shows that:
(1) Most baselines degrade from low to high across all factors, especially for question uncertainty and value grounding complexity (Fig.~\ref{fig:qside_schema} and Fig.~\ref{fig:qside_elements}), because these settings make it harder to identify the relevant structural elements.
(2) \ourmethod{} remains the most robust, achieving the highest accuracy in both groups with smaller drops as question-side complexity increases, demonstrating its effectiveness in handling question uncertainty and complex intents, and identifying relevant structural elements.

\subsection{QA Efficiency (Q3)}
We first compare time efficiency using per-instance answer latency, excluding offline representation construction for representation-based methods and \ourmethod{} (\update{\ourmethod{} takes 14 seconds per table offline}).
Since all baselines are LLM-based, we measure token efficiency by reporting input and output tokens under the GPT-4.1 tokenizer for consistency.
\ifcameraready
\else
Our component-level breakdown is in Appendix \S\ref{appendix_exp_llm_comparison}.
\fi


\begin{figure}[t]
  \centering
  \includegraphics[width=0.7\columnwidth]{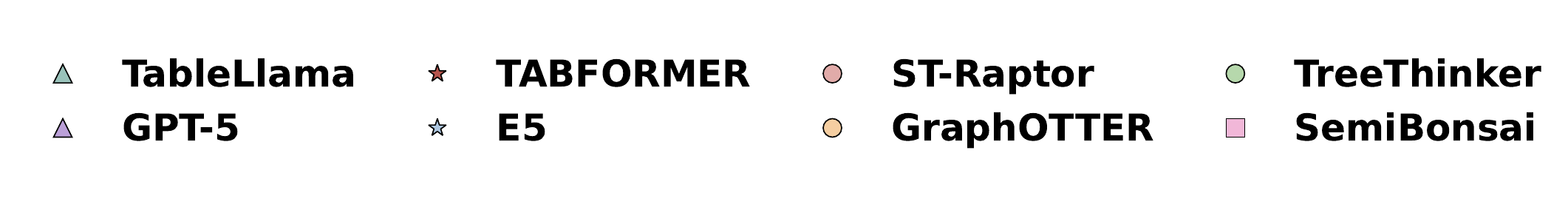}

  \begin{subfigure}[t]{0.32\linewidth}
    \centering
    \includegraphics[width=\linewidth,trim=9mm 10mm 0mm 8mm]{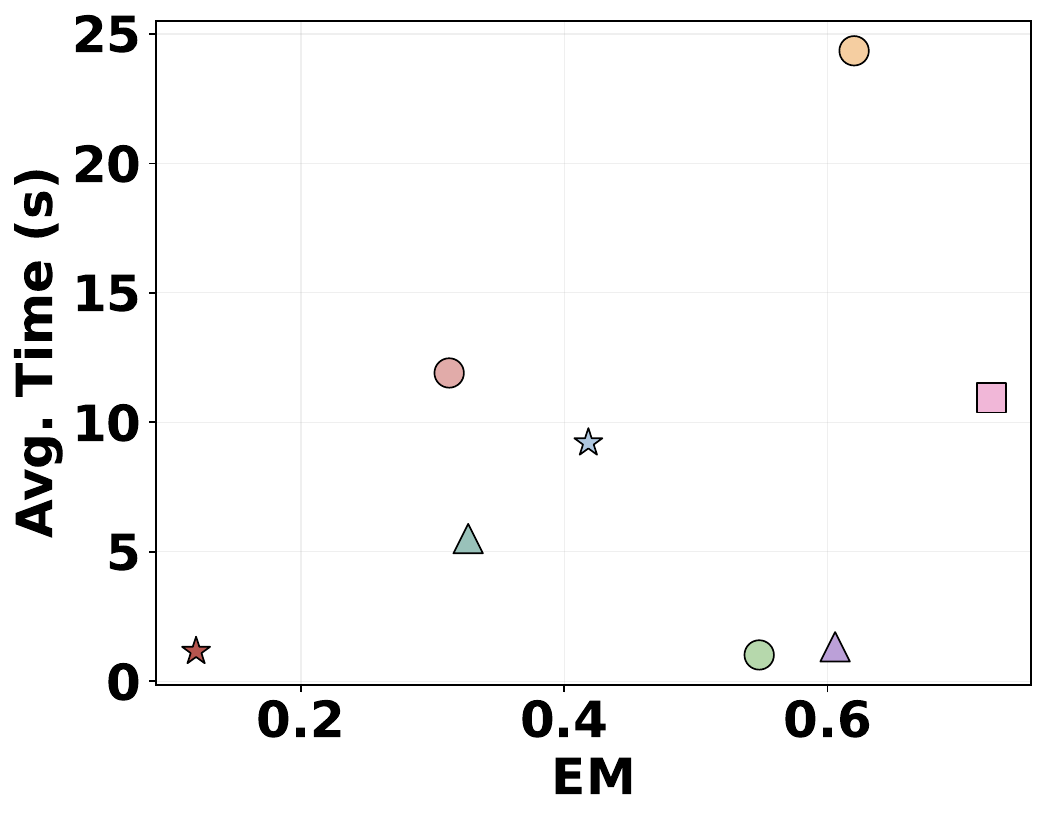}
    \caption{\scriptsize HiTab-Num.}
    \label{fig:tab_tradeoff_hitabnum}
  \end{subfigure}
  \hfill
  \begin{subfigure}[t]{0.32\linewidth}
    \centering
    \includegraphics[width=\linewidth,trim=9mm 10mm 0mm 8mm]{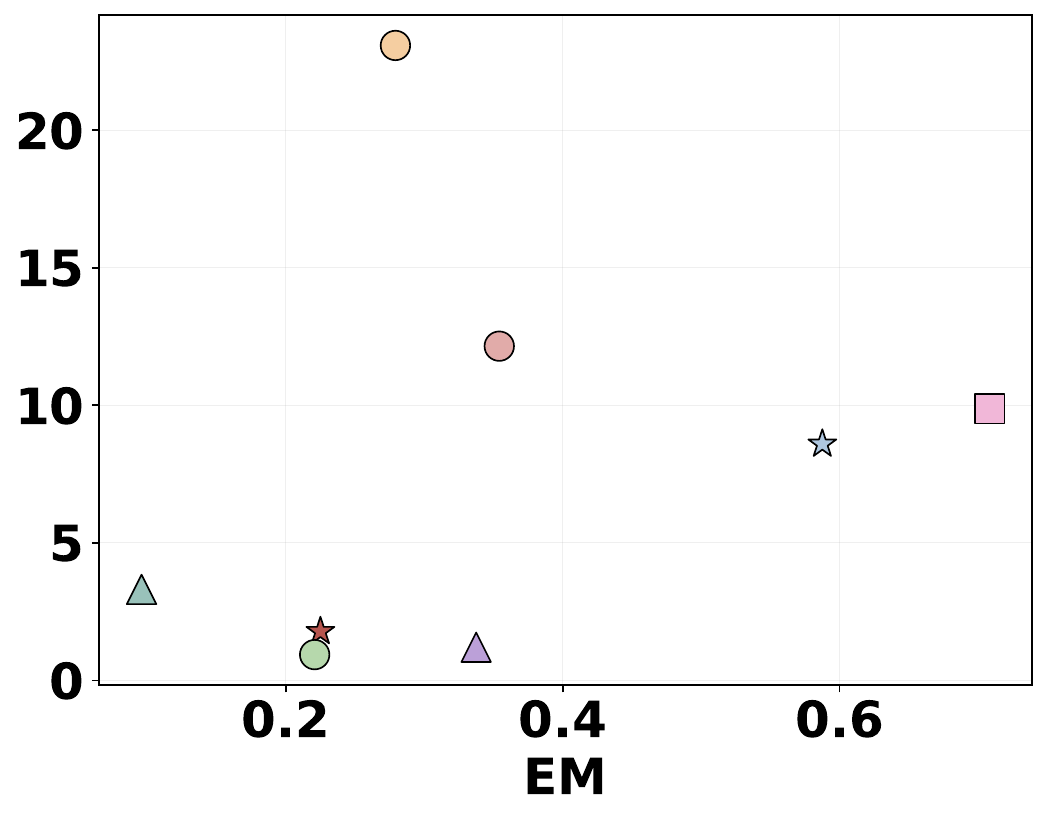}
    \caption{\scriptsize MultiHiertt-Num.}
    \label{fig:tab_tradeoff_mhnum}
  \end{subfigure}
  \hfill
  \begin{subfigure}[t]{0.32\linewidth}
    \centering
    \includegraphics[width=\linewidth,trim=9mm 10mm 0mm 8mm]{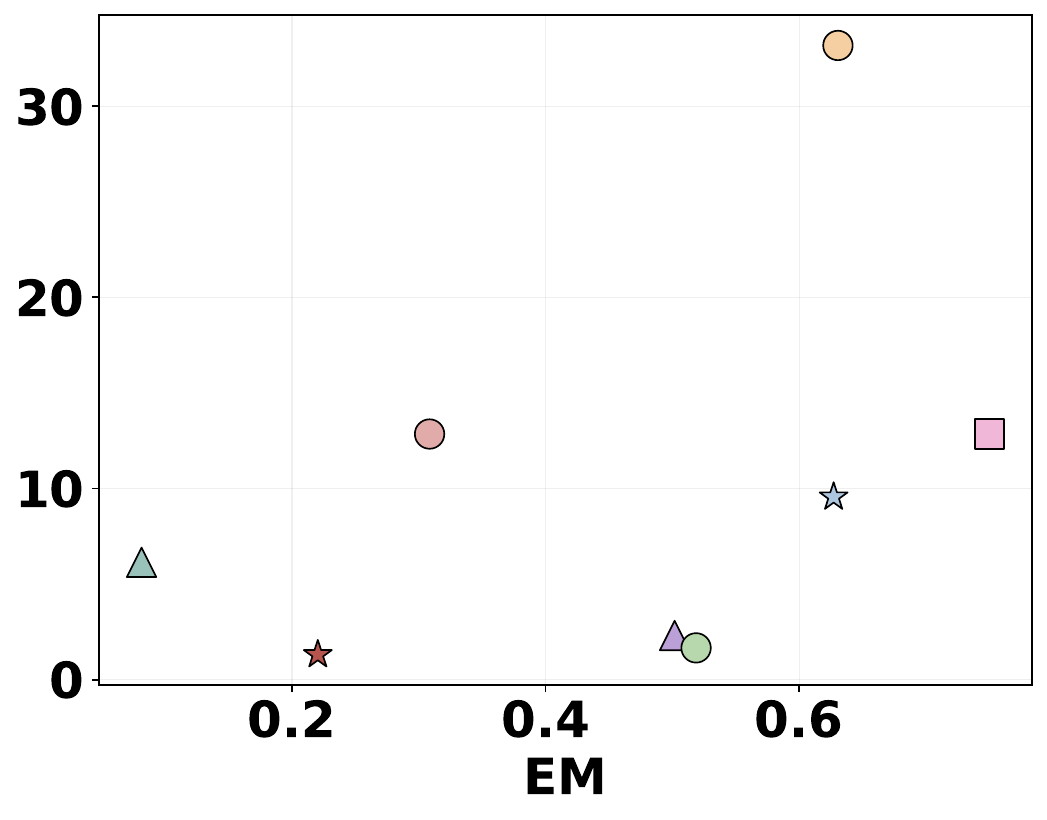}
    \caption{\scriptsize RealHiTBench-Num.}
    \label{fig:tab_tradeoff_rhtnum}
  \end{subfigure}
  \captionsetup{skip=0pt}
  \caption{Per-instance runtime, shown together with answer accuracy to illustrate the trade-off.}
  \label{fig:runtime_compare}
  \vspace{-3pt}
\end{figure}

\begin{figure}[t]
  \centering
  \includegraphics[width=0.7\columnwidth]{figures/complexity_factor_analysis/global_legend_methods.pdf}

  \begin{subfigure}[t]{0.48\linewidth}
    \centering
    \includegraphics[width=\linewidth,trim=9mm 10mm 0mm 8mm]{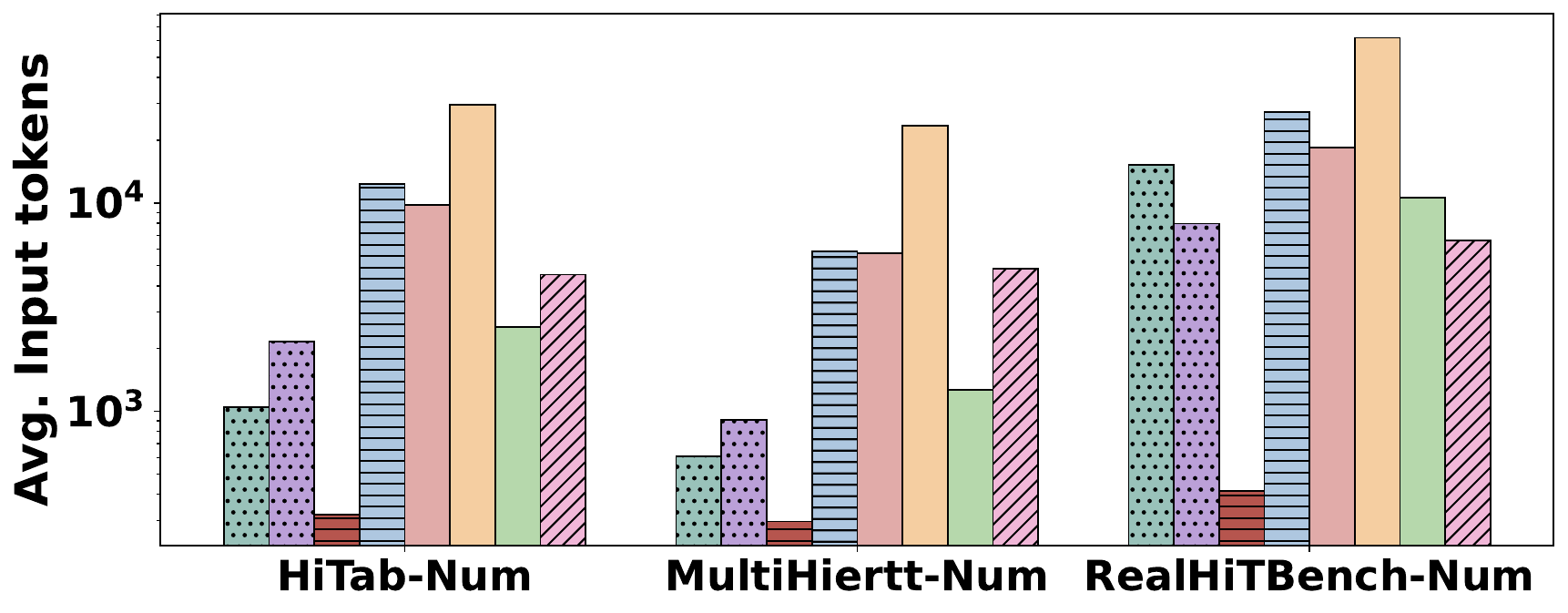}
    \caption{\scriptsize Input tokens per instance}
    \label{fig:token_usage_input}
  \end{subfigure}
  \hfill
  \begin{subfigure}[t]{0.48\linewidth}
    \centering
    \includegraphics[width=\linewidth,trim=9mm 10mm 0mm 8mm]{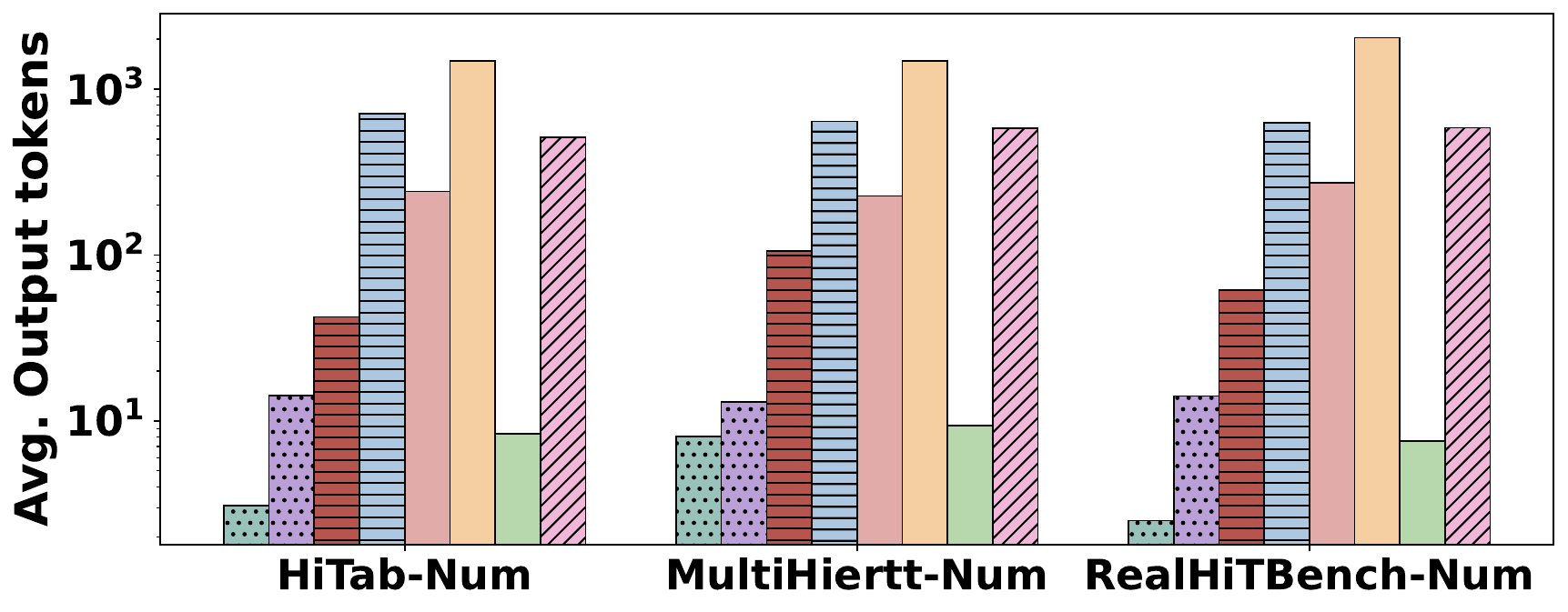}
    \caption{\scriptsize Output tokens per instance}
    \label{fig:token_usage_output}
  \end{subfigure}
  \captionsetup{skip=0pt}
  \caption{Input and output token usage comparison.}
  \label{fig:token_usage}
  \vspace{-3pt}
\end{figure}

\subsubsection{Time efficiency}  
Fig.~\ref{fig:runtime_compare} shows that:
(1) Prompt-based methods are the fastest, as they answer with a single LLM inference. GraphOTTER is the slowest because it repeatedly invokes LLMs to retrieve relevant values before producing an answer. Compared with these baselines, \ourmethod{} incurs a modest runtime.
(2) Across all benchmarks, \ourmethod{} offers the best accuracy–latency trade-off.

\subsubsection{Token efficiency}
Fig.~\ref{fig:token_usage_input} compares input token usage.
Prompt-based methods use the fewest tokens because they mainly input the question and a linearized table, though their usage increases on RealHiTBench-Num due to larger tables (avg.\ 35 rows and 14 columns).
Other baselines show more varied usage: TABFORMER is minimal because it uses only the normalized schema and question, while the remaining methods with repeated LLM-based intermediate reasoning consume substantially more tokens.
\ourmethod{} remains moderate.
Fig.~\ref{fig:token_usage_output} compares output token usage.
Prompt-based methods and TreeThinker use the fewest tokens since they only output a single value, while other methods consume more tokens due to longer intermediate reasoning outputs.

\subsection{Ablation and Component Analysis (Q4)} \label{exp_ablation_study}

In this subsection, we study the effectiveness contribution of the core QA components in \ourmethod{}.
Since the three components are executed sequentially and removing any one of them would break the end-to-end QA pipeline, we instead ablate the key designs within each component while keeping the remaining pipeline unchanged.

\captionsetup[table]{skip=2pt, belowskip=0pt}
\setlength{\textfloatsep}{3pt}      
\setlength{\floatsep}{3pt}          
\setlength{\intextsep}{3pt}         

\begin{table}[t]
  \centering
  \small
  \setlength{\tabcolsep}{1pt}

  \begin{minipage}{\columnwidth}
    \centering
    \begin{threeparttable}
      \begin{tabular}{lcc cc ccc}
        \toprule
        \multirow{2}{*}{Variant} &
        \multicolumn{2}{c}{HiTab-Num} &
        \multicolumn{2}{c}{MultiHiertt-Num} &
        \multicolumn{3}{c}{RealHiTBench-Num} \\
        \cmidrule(lr){2-3}\cmidrule(lr){4-5}\cmidrule(lr){6-8}
        & Row\tnote{1} & Col\tnote{1} & Row & Col & Subtable\tnote{2} & Row & Col \\
        \midrule
        \textsc{LLM Tree} & 0.830 & 0.716 & 0.783 & 0.746 & --\tnote{3} & 0.488 & 0.404 \\
        \textsc{VLM Tree} & 0.823 & 0.822 & 0.792 & 0.869 & 0.649 & 0.467 & 0.646 \\
        \midrule
        \ourmethod{} & \textbf{0.903} & \textbf{0.824} & \textbf{0.898} & \textbf{0.927} & \textbf{0.816} & \textbf{0.677} & \textbf{0.716} \\
        \bottomrule
      \end{tabular}

      \begin{tablenotes}[flushleft]
        \footnotesize
        \item[\tnote{1}] ``Row''/``Col'' denote hierarchical row/column headers.
        \item[\tnote{2}] Only RealHiTBench-Num contains multiple subtables.
        \item[\tnote{3}] \textsc{LLM Tree} does not produce subtable titles on RealHiTBench-Num.
      \end{tablenotes}
    \end{threeparttable}

    \captionsetup{type=table}
    \caption{Layered tree reliability using F1.}
    \label{tab:schema_f1_typewise}
  \end{minipage}

\vspace{3pt}
  \begin{minipage}{\columnwidth}
    \centering
    \small
    \begin{tabular}{lccc}
      \toprule
      Variant & HiTab-Num & MultiHiertt-Num & RealHiTBench-Num \\
      \midrule
      \textsc{LLM Tree} & 0.444$^{\scriptsize \downarrow 39\%}$ & 0.326$^{\scriptsize \downarrow 54\%}$ & 0.413$^{\scriptsize \downarrow 45\%}$ \\
      \textsc{VLM Tree} & 0.438$^{\scriptsize \downarrow 40\%}$ & 0.294$^{\scriptsize \downarrow 59\%}$ & 0.457$^{\scriptsize \downarrow 39\%}$ \\
      \textsc{GT Tree} & 0.846 & 0.783 & 0.888 \\
      
      \midrule
      \textsc{No-Resolver} & 0.572$^{\scriptsize \downarrow 21\%}$ & 0.648$^{\scriptsize \downarrow 9\%}$  & 0.735$^{\scriptsize \downarrow 1\%}$ \\
      \textsc{No-Pruning}  & 0.639$^{\scriptsize \downarrow 12\%}$ & 0.690$^{\scriptsize \downarrow 3\%}$  & 0.743$^{\scriptsize \downarrow 0\%}$ \\
      \midrule
      \textsc{No-Plan}     & 0.706$^{\scriptsize \downarrow 3\%}$  & 0.690$^{\scriptsize \downarrow 3\%}$  & 0.743$^{\scriptsize \downarrow 0\%}$ \\
      \midrule
      \ourmethod{} & \textbf{0.724} & \textbf{0.708} & \textbf{0.744} \\
      \bottomrule
    \end{tabular}

    \captionsetup{type=table}
    \caption{Ablation study using EM.}
    \label{tab:ablation_em}
  \end{minipage}

\end{table}

\noindent \textbf{Structurer.}
Since it aims to produce (i) a reliable logical Layered Tree for answering and (ii) a storage-efficient physical Value Index Structure for value lookup, we study:

\noindent\emph{(1) \underline{Layered tree reliability and QA effectiveness.}}
\update{
Considering that existing tree construction approaches \cite{wu2025realhitbench,tang2025st} lack unified structural elements for modeling, 
we adapt them  to our full node set and compare with our rule-guided Tree construction:
\textsc{LLM Tree}, an LLM-only method adapted from TreeThinker; and
\textsc{VLM Tree}, a VLM-only method that uses the rendered image, adapted from ST-Raptor.
We also include \textsc{GT Tree}, an oracle built from ground-truth structural elements.

}

We evaluate Layered Tree reliability by matching the extracted structural elements against annotated ground truth and reporting F1 for each type following \cite{fuocrbench}.
We treat a predicted structural element as a match only under: the full ordered header sequence and the label text exactly match the ground truth. 
Table~\ref{tab:schema_f1_typewise} shows that \ourmethod{} achieves the highest F1 across all element types and datasets, indicating that our rule-guided construction produces more reliable Layered Trees

We further report the answer accuracy when using each tree in the same pipeline. 
Table~\ref{tab:ablation_em} shows that 
\ourmethod{} achieves substantially higher accuracy than \textsc{LLM Tree} and \textsc{VLM Tree}, because its rule-guided construction yields a more reliable Layered Tree, which directly improves QA effectiveness. 
\ourmethod{} also performs comparably to \textsc{GT Tree}; the remaining gap mainly comes from errors in VLM-based candidate node identification.

\noindent\emph{(2) \underline{Value index structure storage cost.}}
To validate the benefit of our value index structure, we compare with \textsc{Raw Index} (without de-duplication).
On HiTab-Num, MultiHiertt-Num, and RealHiTBench-Num, \textsc{Raw Index} requires 0.9/1.2/5.4 MB, while our Value Index Structure requires only 0.2/0.4/0.8 MB, yielding an average storage reduction of 75\%.

\noindent \textbf{Resolver.}
To quantify the effectiveness of \uncertaintyresolver{}, we compare answer accuracy under:
\textsc{No-Pruning} (disable our context-aware pruning), and
\textsc{No-Resolver} (remove the entire uncertainty resolver module).
Table~\ref{tab:ablation_em} shows that explicitly resolving question uncertainty via the Resolver improves answer accuracy, and that context-aware pruning contributes to these gains via generating candidate groundings and pruning them into a coherent context.
Note that improvement on RealHiTBench-Num is smaller, due to its lower uncertainty (0.31).

\noindent \textbf{Reasoner.} 
To assess the impact of our query plan, we compare it with:
\textsc{No-Plan} (disable query planning).
Table~\ref{tab:ablation_em} shows that incorporating the plan improves accuracy via explicit guidance.
The smaller gain on RealHiTBench-Num is likely due to its fewer complex-intent questions, while on COMPLEX-Q queries with at least 2 operands and 3 operators, the Reasoner achieves a 6\% relative improvement, showing a clearer benefit for complex intents.

\subsection{Cost-effectiveness (Q5)} \label{exp_cost_effectiveness}
\subsubsection{Experimental setup}
To evaluate the cost-effectiveness of \ourmethod{}, we compare against two representative budget-aware LLM routing baselines, FORC~\cite{vsakota2024fly} and PORT~\cite{wu2025port}.
FORC trains a meta-model to predict instance-level accuracy and cost, and then applies ILP to optimize routing decisions.
PORT estimates per-instance answer accuracy and cost by averaging the retrieved similar instances, and then uses an online optimization algorithm for routing.
Although methods such as FrugalGPT \cite{chen2024tmlr} could be adapted to handle budget constraints, they focus on LLM ensembling—invoking multiple LLMs per instance—which differs from our single routing objective; we therefore omit it from comparison.
We also report \textsc{NO Routing}, which always uses the default large LLM.

\noindent\textbf{Evaluation instances, LLM pool and budget constraints.}
We evaluate on three semi-structured table QA benchmarks.
To simulate limited training data while ensuring enough instances for evaluation, we randomly sample 20\% of instances for training and use the remaining 80\% for evaluation on each benchmark. We use an LLM pool consisting of the default LLM GPT-4.1 and six smaller LLMs, including GPT$-4/5$ mini variants and Gemini 2.5-flash variants~\cite{geminimini}. We construct 10 budget constraints by linearly spacing between the minimum and maximum total costs of using a single LLM.

\noindent\textbf{Implementation.}
For \ourmethod{}, we normalize costs to be on the same scale as accuracy, set the contextual bandit exploration parameter to $0.8$ with ridge regularization $1.0$; for the dual update, the initialize cost penalty is $0$ and $\eta$ is $1$.
For the baselines, we use default hyperparameters and set PORT’s retrieval size as $5$ given the small training set.
Since these baselines may exceed the budget, we evaluate all methods under a budget-stop protocol: we execute selections sequentially and stop before the next selection would violate the total budget, and report the accuracy over the evaluation set.
Considering ordering sensitivity, we repeat each constraint for 5 runs with different permutations and report average.

\begin{figure}[t]
  \centering
  \includegraphics[width=0.6\columnwidth]{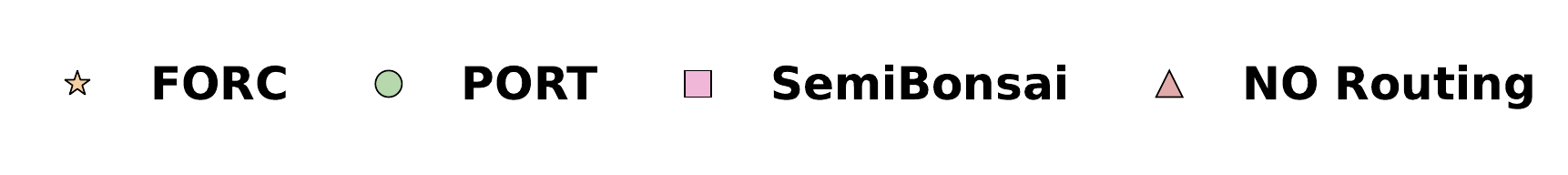}

  \begin{subfigure}[t]{0.32\linewidth}
    \centering
    \includegraphics[width=\linewidth,trim=9mm 10mm 0mm 8mm]{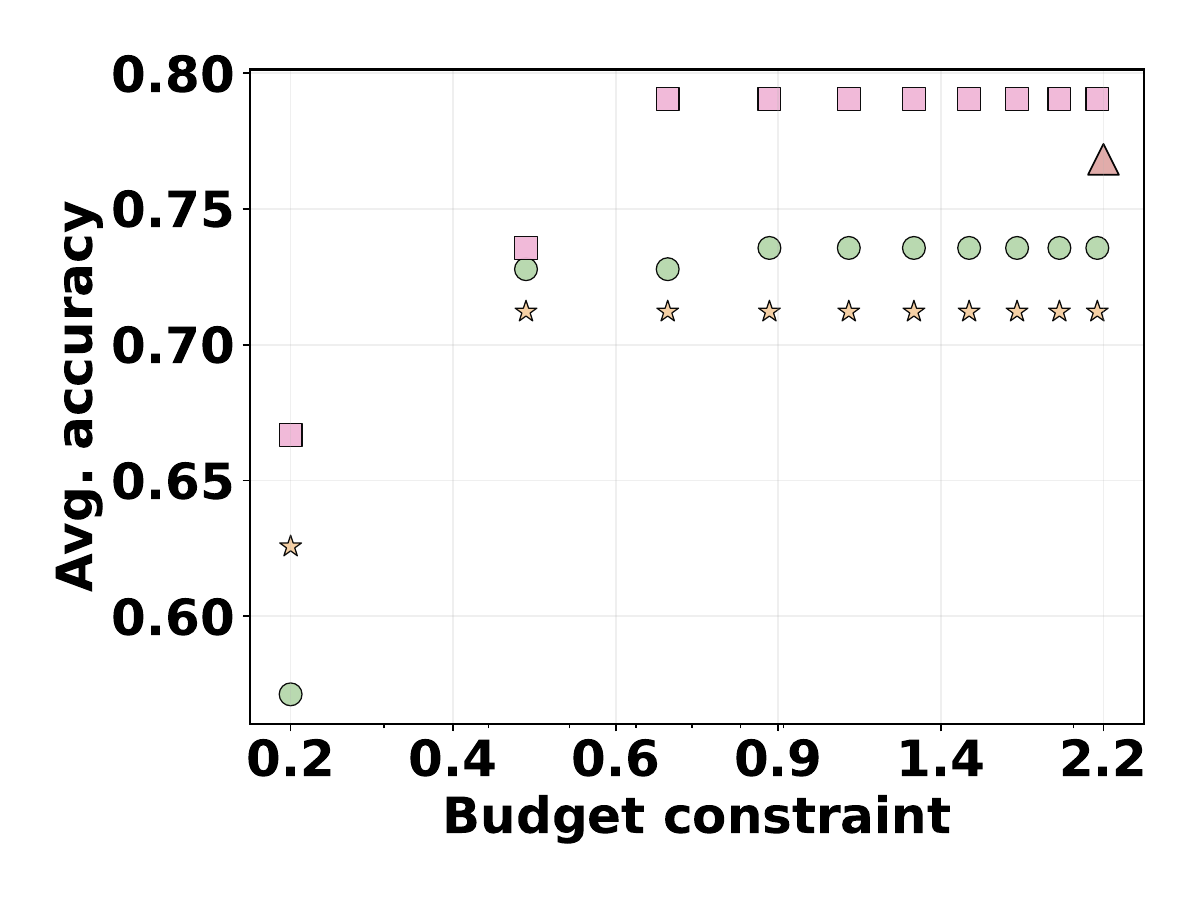}
    \caption{\scriptsize HiTab-Num.}
    \label{fig:costeff_hitabnum}
  \end{subfigure}
  \hfill
  \begin{subfigure}[t]{0.32\linewidth}
    \centering
    \includegraphics[width=\linewidth,trim=9mm 10mm 0mm 8mm]{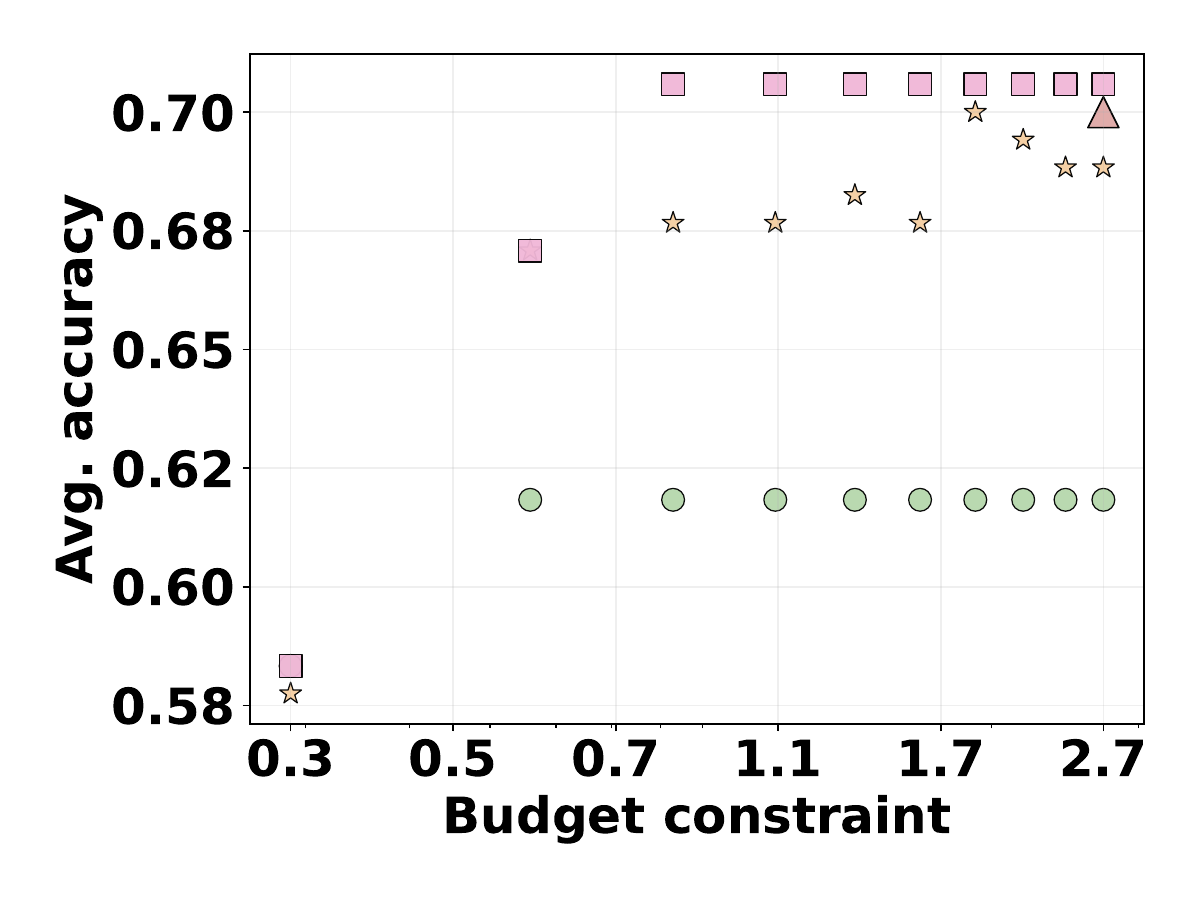}
    \caption{\scriptsize MultiHiertt-Num.}
    \label{fig:costeff_multihierttnum}
  \end{subfigure} 
  \hfill
  \begin{subfigure}[t]{0.32\linewidth}
    \centering
    \includegraphics[width=\linewidth,trim=9mm 10mm 0mm 8mm]{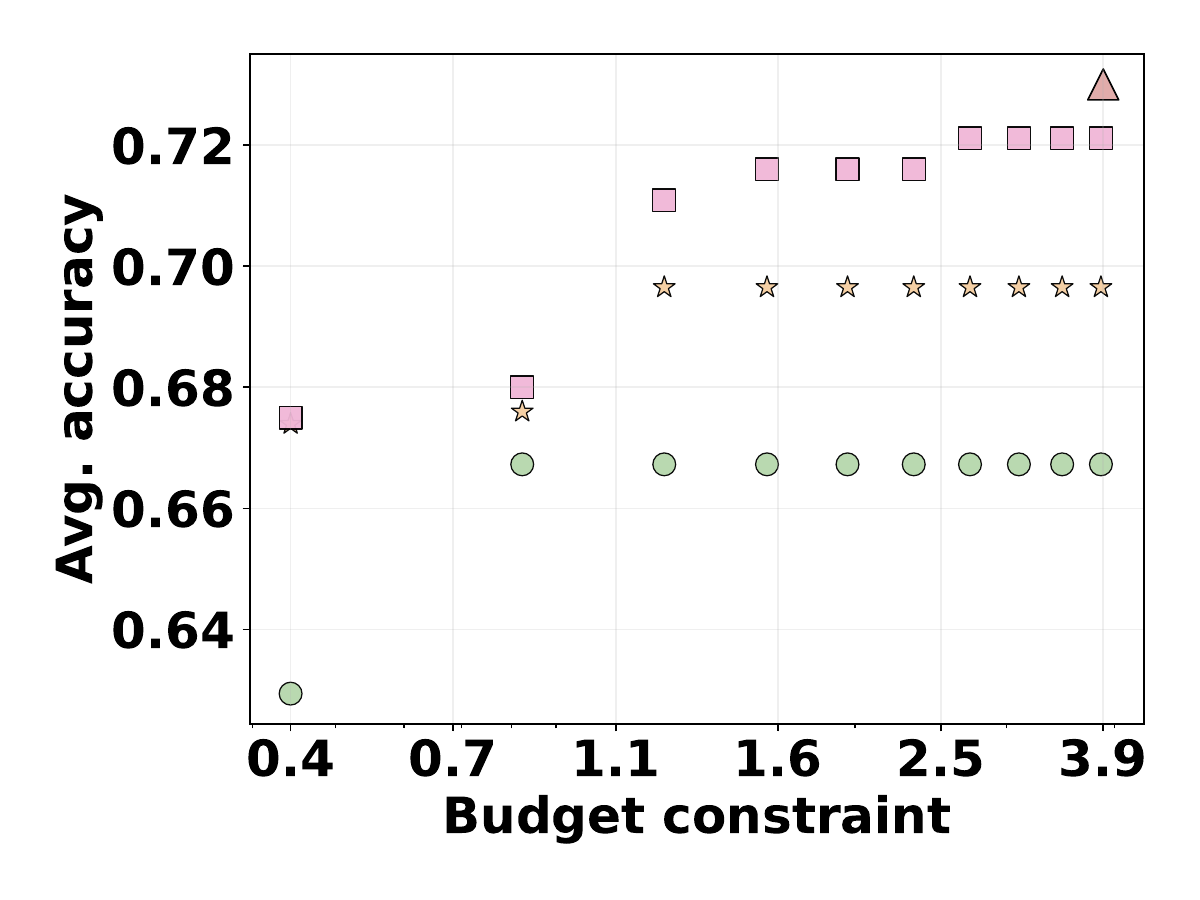}
    \caption{\scriptsize RealHiTBench-Num.}
    \label{fig:costeff_realhitbench}
  \end{subfigure}
  \captionsetup{skip=0pt}
  \caption{Cost--effectiveness: average accuracy achieved under different budget constraints.}
  \label{fig:cost_effectiveness_budget}
\end{figure}
\subsubsection{Main Results} 
Fig.~\ref{fig:cost_effectiveness_budget} reports the average accuracy under different monetary budget constraints.
Observing:
(1) Across all three benchmarks and budgets, \ourmethod{} attains the highest accuracy, compared with routing baselines.
This highlights the effectiveness of our \router{}, which learns per-instance utility for each LLM and adapts selections with budget control, making it more robust with limited training data.
(2) As the budget constraint becomes looser, \ourmethod{} steadily increases accuracy and remains stable. This indicates it can effectively trade cost for answer accuracy, by adaptively updating the cost penalty and route to stronger LLMs when the budget allows.
(3) \ourmethod{} can even surpass \textsc{No Routing}  (Figs.~\ref{fig:costeff_hitabnum} and~\ref{fig:costeff_multihierttnum}).
This is because some smaller LLMs outperform the large LLM on certain instances.
By learning per-instance utility, router identifies these cases and routes them to better-suited models.


\section{Conclusion}
In this paper, we propose \ourmethod{}, a cost-effective, LLM-powered framework 
designed for numerical QA over semi-structured tables. 
\ourmethod{} consists of: 
Multiway Layered Table Structurer, which converts a table into a multiway layered tree;
Context-Aware Uncertainty Resolver, which grounds underspecified phrases to context-coherent structural elements and rewrites the question;
Plan–Guided Reasoner, which decomposes complex intent into a query plan and generates an executable reasoning program; 
Budget-Aware Bandit Router, which learns per-instance routing utilities and applies a budget-constrained control to maximize accuracy under a fixed budget.
Experiments on three benchmarks show that \ourmethod{} consistently improves answer accuracy and achieves strong cost-effectiveness.


\bibliographystyle{IEEEtran}
\bibliography{refer}

\clearpage
\appendices 

\renewcommand{\thesubsection}{\thesection.\arabic{subsection}}
\section{Illustrative Structural Understanding Errors}\label{appendix_structural_error_patterns}
\begin{figure*}[t]
  \centering
  \includegraphics[width=0.85\linewidth]{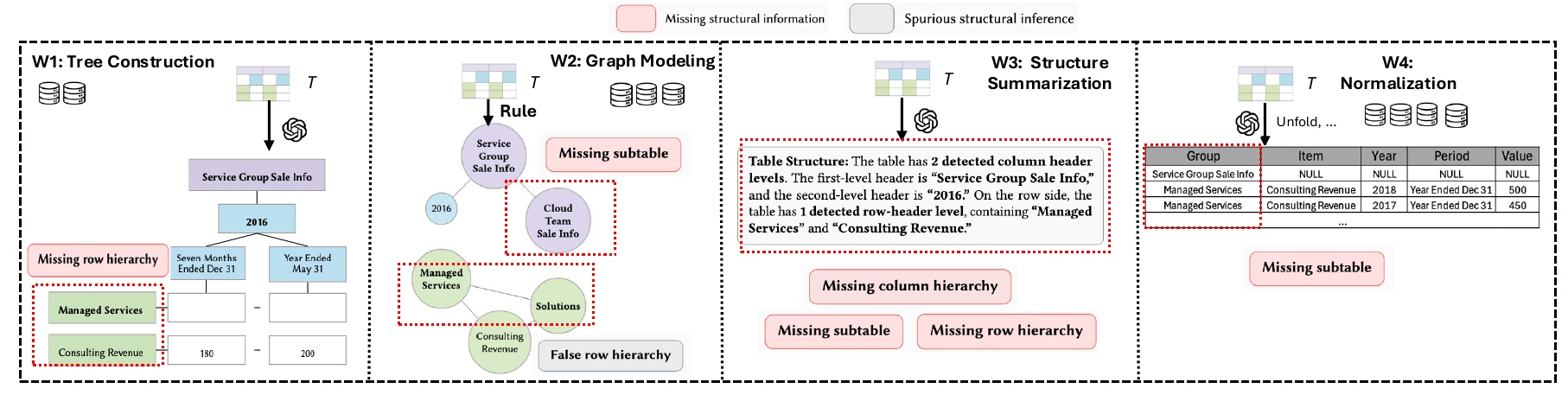}
  \captionsetup{skip=0pt}
  \caption{Error patterns in table structure understanding for representative strategies \textbf{W1}--\textbf{W4}. Required external storage is indicated by \storageicon\ count; \ourmethod{} uses one \storageicon.}
  \label{fig:other_resp}
  \vspace{-3pt}
\end{figure*}

Fig.~\ref{fig:other_resp} illustrates how existing strategies 
\textbf{W1}-\textbf{W4} fail to correctly interpret the 
structural semantics of the running example table $T$ 
introduced in Fig.~\ref{fig:our_resp}. 
\textbf{W1} misses the row hierarchy between \emph{Managed Services} and \emph{Consulting Revenue}; \textbf{W2} mixes the two subtables by linking \emph{Service Group Sale Info} and \emph{Cloud Team Sale Info} and hallucinates an incorrect row hierarchy (e.g., \emph{Managed Services}--\emph{Solutions}); \textbf{W3} outputs a coarse summary that misses subtables and hierarchical row/column header; \update{\textbf{W4} loses the distinct subtable contexts (e.g., \emph{Service Group Sale Info}).
}

\section{Complexity of Questions and Tables} \label{appendix_complex_subset_construction}
As discussed in \S\ref{sec_intro}, the effectiveness depends on the complexity of questions and tables.
To assess this,  we curate three subsets: \(\qsubset{}\) (complex-question subset), \(\tsubset{}\) (complex-table subset), and \(\qtsubset{}\) (both complex question and complex table), by automatically classifying questions and tables into simple or complex.
A straightforward approach is to define a heuristic complexity score using the relevant factors and apply a fixed threshold.
However, the score is sensitive to how multiple factors (e.g., the number and nesting depths of structural elements) interact, and a single threshold may misclassify borderline cases.
Therefore, we 
train feature-based classifiers to learn a data-driven decision boundary that accounts for multiple factors.
Specifically, we (i) construct table-side and question-side complexity features, (ii) build small labeled pools manually, and (iii) train two binary classifiers (for table-side and question-side complexity), which are then applied to all benchmarks to form \(\qsubset{}\), \(\tsubset{}\), and \(\qtsubset{}\).

\subsection{Feature Construction}
\noindent\textbf{Table-side features.}
We extract table-side features using the annotated structural elements provided by each benchmark. Specifically, we compute:
(i) the number of subtables,
(ii) the number of hierarchical row headers and column headers, and
(iii) the statistics (min/mean/max) of their nested depths.

\smallskip
\noindent\textbf{Question-side features.}
We characterize question complexity based on:
(i) \emph{Question uncertainty.} We compute an under-specified ratio, defined as the fraction of key phrases not exactly match the annotated structural elements. Since only HiTab-Num provides key phrase annotations, for the other datasets, we use an LLM to extract key phrases grounded on the table structural elements and then compute the ratio.
(ii) \emph{Complex query intent.} We compute the number of operators and operands, using the annotated reasoning programs and steps.
Since each operator or operand may require traversing multiple structural elements to locate the required values, introducing additional complexity, we compute the number of involved subtables, hierarchical row headers, and hierarchical column headers, for the question, along with their nesting statistics (min/mean/max).

\subsection{Complexity Subset Construction}
\noindent\textbf{Constructing labeled pools.}
We build two small labeled pools (simple vs.\ complex) for tables (69/56) and questions  (139/142) separately as follows. 

\noindent\textit{Labeling tables.} We define a heuristic table complexity score as a weighted sum of two core Table-side features following \cite{tang2025st}: (a) the number of subtables and (b) the mean nesting level of hierarchical row/column headers, with each feature z-score normalized. We rank tables by this score and take the top 20\% as complex and the bottom 20\% as simple. We then manually filter obvious false positives/negatives (e.g., cases where nearly all QA baselines succeed on ``complex'' candidates, or fail on ``simple'' candidates).

\noindent\textit{Labeling questions.} Similarly, we define a heuristic question complexity score as a weighted sum of two core Question-side features: (a)  underspecified ratio and (b) the total number of operands and operators, again with z-score normalization. 
We rank questions by this score, take the top/bottom 20\% as complex/simple candidates, and apply same manual filtering.

\smallskip
\noindent\textbf{Training classifiers and obtaining subsets.}
Using the labeled pools and all constructed features, we train two XGBoost \cite{chen2016corr} binary classifiers—one to predict whether a table is complex and the other to predict whether a question is complex. 
Leveraging all constructed features allows the classifiers to capture diverse complexity factors beyond the heuristic score and yields more reliable classification results. We then apply the classifiers to all examples in the three semi-structured table qa datasets to obtain complex questions and tables.

\section{Lagrangian Relaxation for Routing}
\label{app:routing_theory}
\noindent \textbf{Lagrangian relaxation.}
We start from the budget-aware routing objective (Definition~\ref{sec6_costeffective_obj}), which can be written as the constrained maximization:
\begin{align}
\max_{\pi:\mathcal{D}\to\mathcal{L}} \quad 
& \frac{1}{N}\sum_{k=1}^{N} a\!\left(\pi(q_k,T_k),\,q_k,\,T_k\right) 
\label{eq:app_primal_obj}\\
\text{s.t.}\quad 
& \sum_{k=1}^{N} c\!\left(\pi(q_k,T_k),\,q_k,\,T_k\right) \le B.
\label{eq:app_primal_budget}
\end{align}
Because the policy \(\pi\) makes a discrete choice for each instance, directly solving
\eqref{eq:app_primal_obj}--\eqref{eq:app_primal_budget} is generally hard. We therefore apply a
\emph{Lagrangian relaxation} \cite{fisher1981lagrangian} by introducing a nonnegative multiplier \(\lambda\ge 0\) for the budget
constraint and forming the Lagrangian:
{\footnotesize
\begin{align}
\mathcal{L}(\pi,\lambda)
&= \frac{1}{N}\sum_{k=1}^{N} a\!\left(\pi(q_k,T_k),\,q_k,\,T_k\right)
-\lambda\!\left(\sum_{k=1}^{N} c\!\left(\pi(q_k,T_k),\,q_k,\,T_k\right)-B\right) \nonumber\\
&= \lambda B \;+\; \sum_{k=1}^{N}\left(\frac{1}{N}a\!\left(\pi(q_k,T_k),\,q_k,\,T_k\right) - \lambda\,c\!\left(\pi(q_k,T_k),\,q_k,\,T_k\right)\right).
\label{eq:app_lagrangian}
\end{align}
}

For a fixed \(\lambda\), the term \(\lambda B\) is constant with respect to \(\pi\). Hence maximizing
\(\mathcal{L}(\pi,\lambda)\) over \(\pi\) decomposes across instances, yielding an \emph{instance-wise}
decision rule. Specifically, for each instance \((q_k,T_k)\), the optimal choice under the relaxed
objective is
\begin{align}
\pi_\lambda(q_k,T_k)
&\in \arg\max_{\ell\in\mathcal{L}}
\left(\frac{1}{N}a(\ell,q_k,T_k)-\lambda\,c(\ell,q_k,T_k)\right).
\label{eq:app_dec_rule_with_1overN}
\end{align}

\noindent \textbf{Unconstrained Lagrangian Surrogate Objective.}
Since multiplying the objective by a positive constant does not change the \(\arg\max\), we can absorb
the factor \(1/N\) into the penalty weight by defining \(\tilde{\lambda}=\lambda N\). This yields the
per-instance \emph{utility}:
\begin{align}
\pi_{\tilde{\lambda}}(q_k,T_k)
\in \arg\max_{\ell\in\mathcal{L}}
\underbrace{\left(a(\ell,q_k,T_k)-\tilde{\lambda}\,c(\ell,q_k,T_k)\right)}_{u_{\tilde{\lambda}}(\ell;q_k,T_k)}.
\label{eq:app_utility_rule}
\end{align}

\section{Experiment Supplements}  \label{appendix_exp_llm_comparison}
\begin{table}[t]
\centering
\small
\setlength{\tabcolsep}{3pt}
\begin{tabular}{lccc}
\toprule
Method & \qsubset{} & \tsubset{} & \qtsubset{} \\
\midrule
\multicolumn{4}{c}{\textit{Answer accuracy using EM}} \\
GPT-5.5 & 0.617 & 0.698 & 0.666 \\
\ourmethod{} & \textbf{0.646} & 0.693 & \textbf{0.714} \\
\midrule
\multicolumn{4}{c}{\textit{Monetary cost per instance}} \\
GPT-5.5 & 0.020 & 0.018 & 0.017 \\
\ourmethod{} & 0.015 & 0.017 & 0.017 \\
\bottomrule
\end{tabular}
\caption{Comparison with GPT-5.5 on complex subsets.}
\label{tab:gpt55_complex_comparison}
\end{table}

\update{
\subsection{Comparison with GPT-5.5}
Given the recent release of GPT-5.5~\cite{gpt55}, we further examine whether a stronger frontier LLM can close the performance gap in complex semi-structured table QA.
To this end, we compare \ourmethod{} with GPT-5.5 on the complex subsets.
Table~\ref{tab:gpt55_complex_comparison} presents the results.
Although GPT-5.5 serves as a strong general-purpose LLM baseline, \ourmethod{} achieves higher answer accuracy on \qsubset{} and \qtsubset{}, and comparable EM on \tsubset{}.
Moreover, \ourmethod{} incurs lower or comparable monetary cost per instance across all three subsets, demonstrating a favorable accuracy--cost trade-off under complex settings.
}

\begin{figure}[t]
  \centering
  \begin{subfigure}[t]{0.32\linewidth}
    \centering
    \includegraphics[width=\linewidth,trim=9mm 10mm 0mm 8mm]{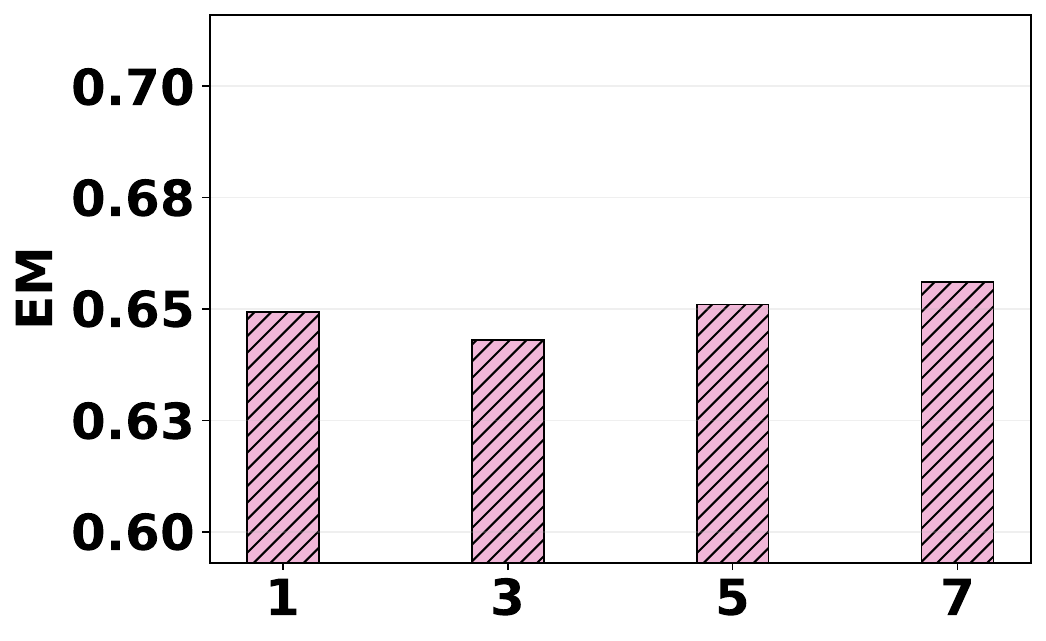}
    \caption{\scriptsize Subtable.}
    \label{fig:aug_subtable}
  \end{subfigure}
  \hfill
  \begin{subfigure}[t]{0.32\linewidth}
    \centering
    \includegraphics[width=\linewidth,trim=9mm 10mm 0mm 8mm]{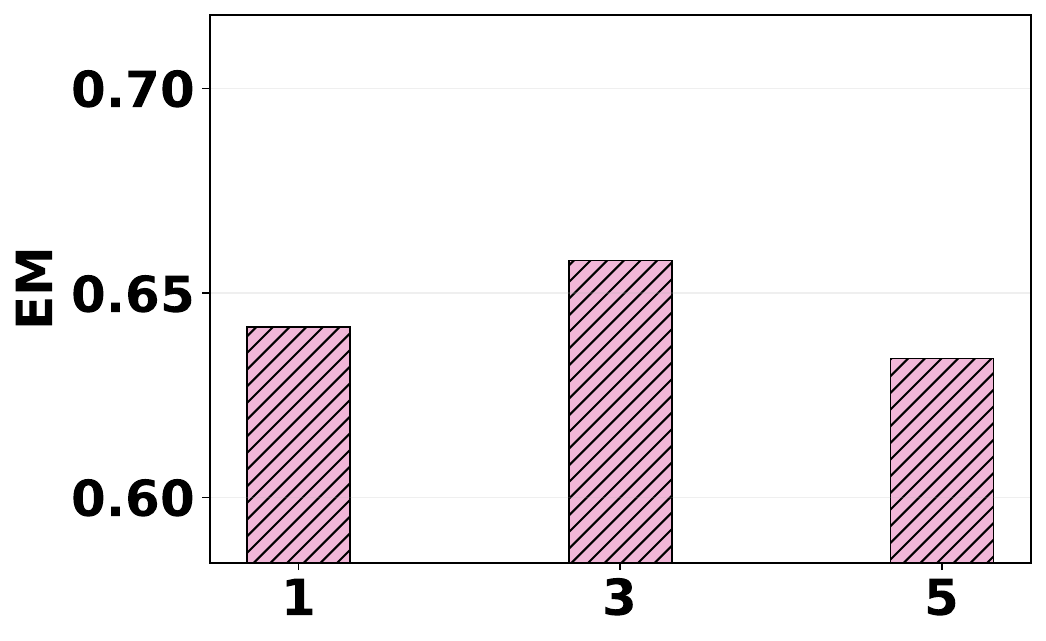}
    \caption{\scriptsize Column header depth.}
    \label{fig:aug_colhdr}
  \end{subfigure}
  \hfill
  \begin{subfigure}[t]{0.32\linewidth}
    \centering
    \includegraphics[width=\linewidth,trim=9mm 10mm 0mm 8mm]{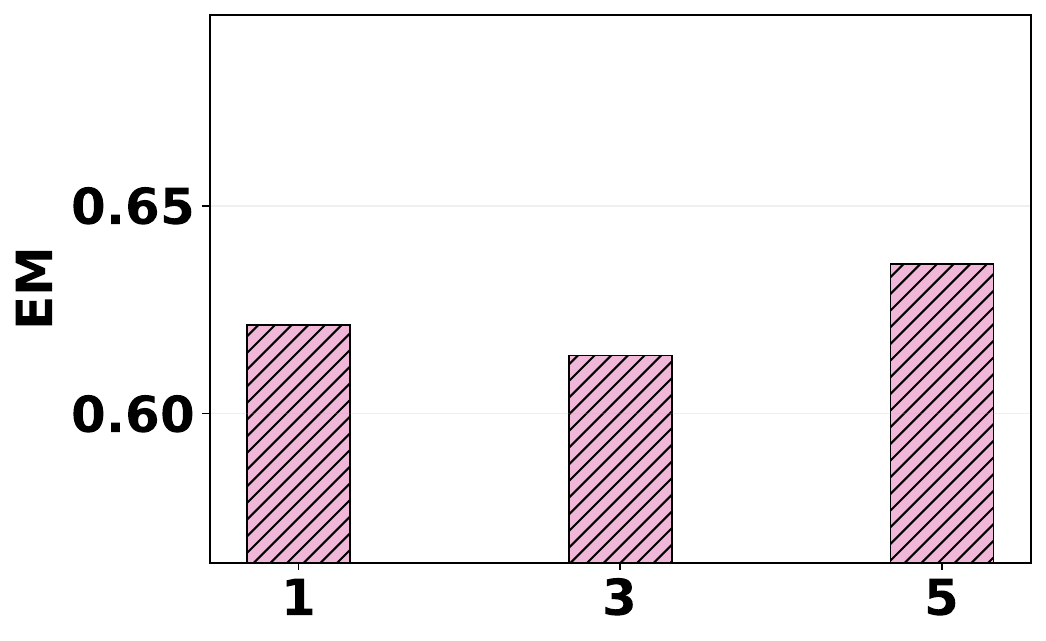}
    \caption{\scriptsize Row header depth.}
    \label{fig:aug_rowhdr}
  \end{subfigure}
  \captionsetup{skip=0pt}
  \caption{Robustness of \ourmethod{} under table-side augmentation on \(\qsubset{}\).}
  \label{fig:table_side_augmented_tables}
\end{figure}
\subsection{Robustness under Complexity-augmented Tables}\label{app:complexity_augmentation}
Considering that existing benchmarks contain few extremely complex tables (e.g., header depths $>3$), we build a complexity-augmented subset by reusing the original QA instances in \(\qsubset{}\) and only increasing table-side complexity while keeping question answer annotations unchanged.

Specifically, for each factor, we first select the lowest-complexity tables (e.g., with a single subtable or depth =1), then prompt an LLM to reorganize the same table content to meet a target complexity level (e.g., increasing header nesting depth or adding more subtables) based on the observed structural elements, without altering the values. We control complexity with discrete targets: subtables \{3,5,7\} and row/column header depth \{3,5\}, chosen to cover complex settings while matching the observed upper bounds in the benchmarks (i.e., max subtables is 7  and depth is 5).

This produces three complexity-augmented collections (subtable, row-header, and column-header augmentation), where each original QA instance is paired with multiple synthesized table variants at different complexity levels.
At each target level, these include 381 questions with 192 subtable variants, 228 questions with 103 row-header variants, and 82 questions with 52 column-header variants.

\noindent\textbf{Results.} As shown in Fig.~\ref{fig:table_side_augmented_tables}, \ourmethod{} remains stable under table-side augmentation as we increase the number of subtables and deepen row/column header hierarchies.
This indicates that its structural model is robust and continues to support reliable downstream answering.

\begin{table}[t]
  \centering
  \footnotesize
  \setlength{\tabcolsep}{1pt}
  \begin{threeparttable}
  \begin{tabular}{lccc ccc ccc}
    \toprule
    Module &
    \multicolumn{3}{c}{HiTab-Num} &
    \multicolumn{3}{c}{MultiHiertt-Num} &
    \multicolumn{3}{c}{RealHiTBench-Num} \\
    \cmidrule(lr){2-4}\cmidrule(lr){5-7}\cmidrule(lr){8-10}
    & Time & InTok\tnote{1} & OutTok
    & Time & InTok & OutTok
    & Time & InTok & OutTok \\
    \midrule
    Structurer & 14.2 & 3,969 & 236 & 14.0 & 3,912 & 176 & 13.2 & 3,747 & 329 \\ \midrule
    Resolver   &  2.7 &   736 & 138 &  1.9 &   702 & 133 &  1.4 & 1,215 &  46 \\
    \midrule
    Reasoner &  8.3 & 3,802 & 375 &  8.0 & 4,144 & 450 & 11.4 & 5,404 & 538 \\
    \quad Planning   &  3.8 & 2,500 & 137 &  4.1 & 2,815 & 173 &  4.9 & 3,449 & 161 \\
    \quad Grounding  &  1.8 &   538 &  92 &  1.6 &   509 & 108 &  2.4 &   843 & 145 \\
    \quad Coding     &  2.7 &   764 & 146 &  2.3 &   820 & 170 &  4.1 & 1,112 & 232 \\
    \bottomrule
  \end{tabular}

  \begin{tablenotes}[flushleft]
    \footnotesize
    \item[\tnote{1}] InTok/OutTok: number of input/output tokens.
  \end{tablenotes}

  \caption{Per-module efficiency of SemiBonsai: runtime (seconds) and token usage per instance.}
  \label{tab:module_efficiency_breakdown}
  \end{threeparttable}
\end{table}
\subsection{Component-level Efficiency} \label{appendix_exp_component_efficiency}
In this subsection, we report a per-component breakdown of runtime and token usage.
For Table Structurer, which preprocesses each table into our structural model based on a VLM, we report per-table preprocessing time and token usage.
For Uncertainty Resolver and Reasoner, which operate online, we report per-question runtime and token usage for end-to-end QA.
As Reasoner incorporates planning, grounding, and coding agents, we further provide each breakdown.

The results are shown in Table \ref{tab:module_efficiency_breakdown}.
The Structurer costs the most
because it invokes the VLM to propose nodes from the table image, then scans the full table to confirm the candidate nodes, and finally infers the structural relations.
Although this may be expensive, it is a one-time cost cached over all questions on the same table.
For online QA, the Resolver is lightweight since it only grounds uncertain phrases and rewrites the question, while the Reasoner dominates usage due to query planning by decomposing the intents.

\begin{table}[t]
\centering
\small
\setlength{\tabcolsep}{1pt}
\begin{tabular}{lccc}
\toprule
Variant & HiTab-Num & MultiHiertt-Num & RealHiTBench-Num \\
\midrule
\textsc{w/o Operand} & 0.713 & 0.704 & 0.743 \\
\ourmethod{} & \textbf{0.724} & \textbf{0.708} & \textbf{0.744} \\
\bottomrule
\end{tabular}
\caption{Impact of operand-based decomposition using EM.}
\label{tab:operand_decomposition_ablation}
\end{table}

\update{
\subsection{The Impact of Operand-based Decomposition}
 \label{appendix_exp_ablation}
We study the effect of operand-based decomposition in Query Planning by comparing \ourmethod{} with a variant that performs only operator-based decomposition.
As shown in Table~\ref{tab:operand_decomposition_ablation}, \ourmethod{} achieves consistent improvements across all datasets.
This suggests that explicitly grounding operands during planning helps better specify complex query intents, thereby benefiting numerical questions that require explicit operand grounding.
}

\section{Prompts} \label{appendix_prompts}
The following lists all the prompts in \ourmethod{}, including the candidate node identification prompt (\S \ref{schema_tree_construction}) of \tablestructurer{}, 
uncertainty identification and question rewriting prompts (\S\ref{method_uncertainty_query_rewriting}) of \uncertaintyresolver{}, planning/grounding/coding prompts (\S\ref{method_reasoner}) of \reasoner{}.

\begin{figure*}[t]
    \centering
    \begin{tcolorbox}[title=Candidate Node Identification Prompt, width=\textwidth]
You are required to extract the table's header structure from a TABLE IMAGE and return a JSON object in a specific format.
\begin{itemize}
    \item {JSON Format}:
        \begin{itemize}
            \item \texttt{column\_header\_groups}: Represent column header structure, can be multi-level or flat.
            \item \texttt{row\_header\_groups}: Represent row header structure, can be multi-level or flat.
        \end{itemize}
    \item {Column Headers}:
        \begin{itemize}
            \item Identify leaf headers (aligned with data columns, left-to-right).
            \item Handle spanning/merged headers: Different output formats based on position relative to leaf headers.
            \item Output flat items if no spanning/shared headers.
            \item Do not mix flat and spanning-based items for the same columns.
            \item Use only cell texts.
        \end{itemize}
    \item {Row Headers}:
        \begin{itemize}
            \item Check left-most column below header band.
        \end{itemize}
    \item {Rules}:
        \begin{itemize}
            \item Do not mix columns from different visual blocks.
            \item Use verbatim text from the image.
            \item Output only the specified JSON object.
        \end{itemize}
\end{itemize}
\end{tcolorbox}
\end{figure*}

\begin{figure*}
    \centering
\begin{tcolorbox}[title=Uncertainty Identification and Resolution Prompt,width=\textwidth]
You are a table-question grounding assistant.

You are given:
\begin{itemize}
    \item (1) a question, and
    \item (2) table metadata node labels (unique labels), consisting of:
    \begin{itemize}
        \item \texttt{subtable\_titles}: section/subtable titles
        \item \texttt{column\_headers}: column header node labels
        \item \texttt{row\_headers}: row header node labels
    \end{itemize}
\end{itemize}

Your task has {two parts}:

\begin{itemize}
    \item {A) Identify underspecified query phrases}
    \begin{itemize}
        \item Extract all phrases in the question that are underspecified/ambiguous in the context of the table metadata.
        \item A phrase is underspecified if it cannot be grounded to a specific table element without interpretation, \emph{or} it could refer to multiple metadata nodes.
        \item If there is {no} underspecified phrase, return \texttt{\{"phrase\_groundings": []\}}.
    \end{itemize}

    \item {B) Ground each phrase to relevant metadata node labels (one-to-many)}
    \begin{itemize}
        \item For {each} underspecified phrase, select {all} relevant node labels it could refer to.
        \item Return selections in three buckets: \texttt{subtable\_titles} / \texttt{column\_headers} / \texttt{row\_headers}.
        \item Selections must be {strictly} chosen from the provided metadata lists (no invented labels).
        \item Rank the selected labels within each bucket by confidence (most likely first).
        \item If a bucket has no relevant labels, return an empty list for that bucket.
    \end{itemize}

    \item {Strict rules}
    \begin{itemize}
        \item Use {only} labels that appear in the provided metadata lists (verbatim; no paraphrasing).
        \item Do {not} do any global consistency or cross-phrase pruning (handled later).
        \item Output must be {valid JSON} and nothing else.
    \end{itemize}

    \item {Output JSON format}
\begin{verbatim}
{
  "phrase_groundings": [
    {
      "phrase": "...",
      "selected_nodes": {
        "subtable_titles": [...],
        "column_headers": [...],
        "row_headers": [...]
      }
    }
  ]
}
\end{verbatim}
\end{itemize}

{Question:} \{question\} \\
{Metadata:} \{table\_metadata\}
\end{tcolorbox}
\end{figure*}
\begin{figure*}
    \centering
    \begin{tcolorbox}[title=Question Rewriting Prompt,width=\textwidth]
You are given:
\begin{itemize}
    \item (1) an original question, and
    \item (2) a set of {selected structural elements} that map underspecified phrases.
\end{itemize}

Your task:
\begin{itemize}
    \item Rewrite the question into a clear, fully specified version that explicitly mentions the grounded labels.
    \item Preserve the original intent and semantics.
    \item Do {not} add any new facts not implied by the groundings.
\end{itemize}

{Rules}
\begin{itemize}
    \item Use the grounded labels {verbatim}.
    \item Output must be {valid JSON} and nothing else.
\end{itemize}

{Output JSON format}
\begin{verbatim}
{
  "rewritten_question": "...",
  "used_groundings": [
    {
      "phrase": "...",
      "structural_elements": [...]
    }
  ]
}
\end{verbatim}

{Original Question:} \{question\} \\
{Selected elements:} \{selected\_groundings\}
\end{tcolorbox}
\end{figure*}

\begin{figure*}[t]
\centering
\begin{tcolorbox}[title=Planning Prompt, width=\textwidth]
You are an LLM-based planning agent for numerical table QA.

\textbf{Subintent.} $\kappa=\langle \texttt{func},\texttt{subqs},\texttt{rawq},\texttt{rel}\rangle$:
\texttt{func} (operator/computation), \texttt{subqs} (subquestions for operands/params), \texttt{rawq} (NL), \texttt{rel} (dependencies).

\textbf{Goal.} Pick one unresolved target from \texttt{subquestions} using \texttt{action history}, then output the next action.

\textbf{Allowed actions (pick one).}
\texttt{infer\_calculation\_formula} \quad
\texttt{multihop\_question\_decomposition} \quad
\texttt{generate\_execute\_program}

\textbf{Rules.}
(i) For the first two actions, the first argument must be the target subquestion ID. 
(ii) If decomposing, create new unused subquestion IDs.

\textbf{Decision.}
\begin{itemize}
    \item \textbf{Directly executable} (retrieve cell(s) from metadata, optionally simple aggregation; no formula/hops) $\rightarrow$ \texttt{generate\_execute\_program}.
    \item \textbf{Numerical computation} (operands decomposition) $\rightarrow$ \texttt{infer\_calculation\_formula}:
    infer minimal \texttt{func} using operators in \texttt{\{operator\_set\}} with placeholders \#1,\#2,\dots; create \texttt{subqs} where each new subquestion supplies one operand; record dependencies in \texttt{rel}.
    \item \textbf{Multi-hop reasoning} (operators decomposition) $\rightarrow$ \texttt{multihop\_question\_decomposition}:
    extract the first operator as \texttt{func}, decompose into $\leq 3$ single-hop subquestions (ordered); encode inter-hop dependencies in \texttt{rel}.
\end{itemize}

\textbf{Inputs.} \texttt{subquestions: \{subquestions\}} \quad
\texttt{table\_metadata: \{table\_metadata\}} \quad
\texttt{action\_history: \{operation\_history\}} \quad
\texttt{possible\_actions: \{possible\_actions\}}

\textbf{Output (JSON only).}
\begin{verbatim}
{"function":"...","explanation":"..."}
\end{verbatim}
\end{tcolorbox}
\end{figure*}
\begin{figure*}
\centering
\begin{tcolorbox}[title=Grounding Prompt,width=\textwidth]
You are given a question and table metadata labels:
\texttt{subtable\_titles}, \texttt{column\_headers}, \texttt{row\_headers}.

\textbf{Task.}
(1) Extract key query phrases. 
(2) For each phrase, select all relevant labels from the metadata (one-to-many), grouped into the three buckets above and ranked by confidence.

\textbf{Rules.} Use only provided labels (verbatim). Output valid JSON only.

\textbf{Output (JSON).}
\begin{verbatim}
{"phrase_groundings":[{"phrase":"...","selected_nodes":{
  "subtable_titles":[...],"column_headers":[...],"row_headers":[...]}}]}
\end{verbatim}

\textbf{Question:} \{question\}\quad
\textbf{Metadata:} \{table\_metadata\}
\end{tcolorbox}
\end{figure*}
\begin{figure*}
\centering
\begin{tcolorbox}[title=Coding Prompt,width=\textwidth]
Generate one complete, directly executable {Python} snippet to answer the {raw} numerical question using the given atomic subquestions and relevant subtable data.

\textbf{Critical output rules.}
(i) Output only a single \verb|```python ... ```| block. 
(ii) No comments in code. 
(iii) Assign the final result to \texttt{final\_answer}.

\textbf{Data rules (Pandas).}
Build a minimal DataFrame via \texttt{pd.DataFrame} from a Python dict using only the relevant subtable; include only needed rows/columns; normalize numeric strings (e.g., commas, \%, currency) to \texttt{int}/\texttt{float}; filter out summary rows (e.g., ``Total'', ``Subtotal'') if they affect aggregation.

\textbf{Task.}
Compute intermediate values for atomic subquestions first, then combine them to answer the raw question. Use reasoning hints and program evidence if helpful.

\textbf{Inputs.}
\texttt{atomic\_subquestions: \{atomic\_subquestions\}} \quad
\texttt{relevant\_subtable: \{atomic\_subquestions\_subdata\}} \quad
\texttt{subquestions: \{subquestions\}} \quad
\texttt{reasoning\_history: \{reasoning\_history\}} \quad
\texttt{program\_evidence: \{numerical\_reasoning\_context\}}
\end{tcolorbox}
\end{figure*}

\end{document}